\documentclass[pre,aps,showpacs,preprint]{revtex4}

\usepackage{inputenc}
\inputencoding{ansinew}

\usepackage[T1]{fontenc}

\usepackage{graphicx}

\usepackage[Gray,squaren]{SIunits}
\usepackage{amsmath}
\usepackage{amssymb}
\usepackage{longtable}
\usepackage{pstricks}
\usepackage{float}
\usepackage{setspace,supertabular}
\usepackage{longtable}
  \usepackage{palatino}
  \usepackage{mathpazo}
\usepackage{epsfig} % include EPS-Figures
\usepackage{graphicx}
\usepackage{epsfig}
\usepackage{amsfonts}
\usepackage{amsmath}
\usepackage{bm}
\usepackage{amssymb}
\usepackage{natbib}
\usepackage{color}
\definecolor{MycBlue}{rgb}{0,0,1}
\definecolor{MycGreen}{rgb}{0,1,0}
\definecolor{MycBlack}{rgb}{0,0,0}
\definecolor{MycRed}{rgb}{1,0,0}
\definecolor{MycCyan}{rgb}{0,1,1}
\definecolor{MycMagenda}{rgb}{1,0,1}
\definecolor{MycYell}{rgb}{1 0.333 0 }
\newcommand*{\eps}{\varepsilon}
\newcommand{\am}{\alpha_{\rm min}}
\newcommand{\fm}{f_{\rm max}}
\newcommand{\wh}{w_{1/2}}

\begin{document}
\title{Prevalence of approximate $\sqrt t$ relaxation for the dielectric $\alpha$ process in viscous organic liquids}
\author{Albena I. Nielsen$^a$, Tage Christensen$^a$, Bo Jakobsen$^a$, Kristine Niss$^a$, \\ Niels Boye Olsen$^a$, Ranko Richert$^b$, and Jeppe C. Dyre$^a$}

\affiliation{$^a$DNRF Centre ``Glass and Time'', IMFUFA, Department of Sciences,
Roskilde University, Postbox 260, DK-4000 Roskilde, Denmark\\
$^b$Department of Chemistry and Biochemistry, Arizona State University, Tempe, Arizona 85387-1604, USA}
\date{\today}
\keywords{viscous liquids, structural relaxation, primary relaxation, generic properties, dielectric, minimum slope, time-temperature superposition, width, temperature index}

\begin{abstract}
This paper presents dielectric relaxation data for organic glass-forming liquids compiled from different groups and supplemented by new measurements. The main quantity of interest is the ``minimum slope'' of the $\alpha$ dielectric loss plotted as a function of frequency in a log-log plot, i.e., the numerically largest slope above the loss peak frequency. The data consisting of 347 spectra for 53 liquids show prevalence of minimum slopes close to $-1/2$, corresponding to approximate squareroot(time) dependence of the dielectric relaxation function at short times. The paper studies possible correlations between minimum slopes and: 1) Temperature (quantified via the loss-peak frequency); 2) How well an inverse power law fits data above the loss peak; 3) Degree of time-temperature superposition; 4) Loss-peak half width; 5) Deviation from non-Arrhenius behavior; 6) Loss strength. 
For the first three points we find correlations that show a special status of liquids with minimum slopes close to $-1/2$. For the last three points only fairly insignificant correlations are found, with the exception of  large-loss liquids that have minimum slopes that are numerically significantly larger than $1/2$ and half loss peak widths that are significantly smaller than those of most other liquids. We conclude that -- excluding large-loss liquids -- approximate $\sqrt t$ relaxation appears to be a generic property of the $\alpha$ relaxation of organic glass formers.
\end{abstract}

%\pacs{64.70.P-, 64.70.Dv, 71.15.Pd, 61.50.Ah, 61.66.?f}
\maketitle
%opening

\section{Introduction}

The glass transition takes place when a liquid is cooled so fast that it does not have sufficient time to equilibrate \cite{gut95,edi96,chamberlin99,ang00,don01,dyre06}. Below the glass transition temperature $T_g$ the sample is in a solid but structurally disordered state, where the molecular positions are akin to those of the higher-temperature supercooled liquid state. Above $T_g$ the liquid is in metastable equilibrium, but generally has much longer relaxation time than less-viscous liquids like ambient water. This makes the study of relaxation processes in highly viscous liquids possible and useful for obtaining information about these liquids' dynamical properties.

Physical systems usually relax with time following perturbations forced upon them. The simplest form of relaxation is an exponential decay towards equilibrium. This is, however, rarely observed. Another simple case is the so-called $\sqrt t$ relaxation where the relaxation function $h(t)$ at short times decays as $h(0)-h(t)\propto\sqrt t$. This is observed in systems as diverse as Rouse dynamics of polymer chains \cite{doi86}, metallic glasses \cite{wang2008}, molecular nanomagnets \cite{wer99,wer06}, and turbulent transport, e.g., in astrophysics \cite{bak04}. For random walks, the equivalent of $\sqrt t$ relaxation is referred to as single-file diffusion which is observed, e.g., in ion channels through biological membranes, diffusion in zeolites, and charge-carrier migration in one-dimensional polymers \cite{mon02}. 

Below we present data showing prevalence of $\sqrt t$ relaxation in glass-forming organic liquids. The data were taken on organic liquids studied in the extremely viscous state just above the glass transition where the relaxation time is in some cases larger than 1 second. In a paper from 2001 the equivalent of $\sqrt t$ relaxation -- high-frequency dielectric losses decaying as $\propto f^{-1/2}$ where $f$ is frequency -- was linked to time-temperature superposition (TTS) via the conjecture that the better a liquid obeys TTS, the more accurate is $\sqrt t$ relaxation obeyed \cite{Olsen01}. The present paper takes a slightly different approach by not focusing specifically on possible correlation to TTS, but on the overall behavior of viscous liquids. From a compilation of dielectric relaxation data from leading groups in the field supplemented by own measurements for altogether 53 organic liquids we find a clear prevalence of $\sqrt t$ relaxation. Every effort has been made to avoid possible bias in the data selection. It is important to note, however, that no objective criteria have been applied for choosing the liquids -- they were included whenever data of sufficient quality happened to be available to us. 

Relaxation processes in supercooled liquids occur over a wide range of time scales. The typical processes observed in viscous liquids (e.g., by dielectric relaxation spectroscopy) are the slow, primary, so-called $\alpha$ process that is associated with the calorimetric glass temperature, and the faster secondary \cite{note1} $\beta$ process(es) \cite{note2,JohGold70,JohGold71}. These processes almost always deviate from what corresponds to a simple exponential relaxation function \cite{dyre06, jaeckel86}, a Debye frequency dependence. The relation between $\alpha$ and $\beta$ processes manifests itself differently for different liquids. In many cases they are observed as two separate processes with well-defined and clearly distinguishable relaxation times. In other cases the $\beta$ process is partly hidden by the primary process and manifests itself only as a high-frequency wing \cite{Olsen98, lunkeh,gai2006}. 

The time scales of the $\alpha$ and $\beta$ processes may be separated by lowering temperature or increasing pressure. The $\beta$ process does not slow down significantly on lowering temperature as long as one works in the equilibrium liquid phase \cite{Olsen98, wang2007b}; in some cases it even becomes faster as temperature is lowered \cite{Olsen98,dyr03}. If the $\beta$ process is in the high-frequency end of the experimental window, a clear separation between $\alpha$ and $\beta$ relaxations appears upon cooling. A similar increased separation is observed when pressure is increased at constant temperature because the $\alpha$ process slows down considerably with compression while the $\beta$ relaxation time is almost pressure insensitive \cite{pawlus, capaccioli}. Furthermore, as pressure increases at constant temperature one generally finds that the $\beta$ process' intensity decreases, which reduces its influence on the $\alpha$ process \cite{Olsen98,pawlus, bielowka}.

The $\alpha$ process has a characteristic asymmetry. This is reflected in the popular fitting function, the stretched exponential (Kohlrausch-Williams-Watts, KWW) function $h(t) = h_0 \exp \left[ -\left( t/\tau \right)^ {\beta_{KWW}}\right]$ \cite{kol,wil,williams,cardona}. The parameter $0<\beta_{KWW}<1$ is termed the stretching exponent. An alternative fitting function is the Cole-Davidson (CD) function which relates directly to the frequency domain by predicting for the dielectric constant $\eps(\omega) - \eps_{\infty }= \Delta\eps (1+i \omega \tau)^{-\beta_{CD}}$ \cite{CD1, CD2}.  For both functions, in a log-log plot the slope on the high-frequency side of the dielectric loss (the negative imaginary part of the dielectric constant) converges to $-\beta_{KWW}$ and $-\beta_{CD}$, respectively \cite{note3}. Typical values of these quantities reported in the literature range between $0.3$ and $0.7$ \cite{boehmer1993,wang2007b}. Thus the typical high-frequency decay of the $\alpha$ dielectric loss is somewhere between $\propto f^{-0.3}$ and $\propto f^{-0.7}$ (although there are also several exceptions to this). This is the ``conventional wisdom'' of the field, where no exponent is supposed to be more typical than any other but with a strong correlation with fragility. In contrast to this, we find below a prevalence of what corresponds to $\beta_{KWW}=1/2$ or $\beta_{CD}=1/2$ at high frequencies for liquids covering a wide range of fragilities. We do not fit the data to these two fitting functions, though, but analyze data directly without fitting to particular functions; in fact we find a range of widths at half loss, showing that none of these two functions fit data accurately.

There are reports in the literature of a number of liquids that have power-law exponent close to $-1/2$  \cite{Olsen01,Jacobsen05,wang2007b}. As already mentioned, Olsen et al. in 2001 \cite{Olsen01} conjectured that if the $\alpha$-process obeys time-temperature superposition accurately, the frequency dependence of the high-frequency $\alpha$ loss is close to having the universal exponent $-1/2$, i.e.,

\begin{eqnarray}\nonumber
\eps '' (f) \varpropto f^{-1/2}, \, f>>\fm\,.
\end{eqnarray}
Is this particular exponent predicted by any models? The answer is yes; in fact there are quite a few models predicting a high-frequency exponent of $-1/2$ (see, e.g., Refs. \cite{Dyre05a, Dyre05b} and their references). In the 1960's and 1970's, in particular, several theories were proposed predicting this exponent, famous among which are: Glarum's defect diffusion model \cite{glarum,doremus,bordewijk}; the ``inhomogeneous media'' model of Isakovich and Chaban \cite{isak}; the Barlow-Erginsav-Lamb (BEL) model postulating a mechanical equivalent of a simple electrical circuit \cite{Barlow,Dyre05a}; the Montrose-Litovitz model invoking diffusion and relaxation of some unspecified order \cite{montrose1970}. The idea of a universal exponent equal to $-1/2$ gradually fell out of favor, however, to be replaced by the presently popular view that relaxation functions are basically determined by the fragility \cite{boehmer1993}.

In this work we present an empirical investigation of the best dielectric data we could acquire, resulting in a collection of data for 53 organic glass formers. The data were collected in order to investigate whether or not the exponent $-1/2$ has a particular significance. As mentioned, this exponent for the high-frequency decay of the relaxation function corresponds to $\sqrt t$ relaxation in the time domain. The possible prevalence of exponent $-1/2$ is investigated by analyzing dielectric relaxation, and not other, data. This is because the complex dielectric permittivity is by far the most accurately measured of all relaxing quantities and, furthermore, this quantity is available for many liquids measured over broad frequency ranges \cite{kre02}. Numerous dielectric measurements have been published on different liquids, and dielectric spectroscopy setups continuously improve \cite{kud99,rol05}. In order to make the procedure as objective as possible the data analysis used is model independent and, as far as possible, automated. ``Model independent'' means that data are analyzed in terms of quantities obtained directly from the raw data.

Simple monoalchohols were excluded from the analysis because of their well-known dominant low-frequency Debye-like relaxation that is not related to the calorimetric glass \cite{Huth07}.  Similarly, plastic crystals and polymers were  excluded because their glass transitions are not a liquid-glass transition. Besides this no selection criteria were applied except that too noisy data were discarded.

In Sec. II experimental details are provided and new data are presented. Section III discusses data selection criteria and details of the data analysis. Section IV presents the results for the minimum slopes in the form of a histogram. Section V analyzes various possible correlations by investigating whether minimum slopes correlate with: 1) how well an inverse power-law describes the high-frequency loss, 2) temperature, 3) how well time-temperature superposition applies, 4) loss peak width, 5) deviations from Arrhenius behavior, and 6) dissipation magnitude. Section VI summarizes our findings.

\section{Experimental}

The 53 liquids studied in this paper are listed in Table \ref{tabel} that for each liquid gives temperature and frequency ranges, etc. Part of the data analyzed were kindly provided by the R{\"o}ssler group (Bayreuth, Germany), the Loidl-Lunkenheimer group (Augsburg, Germany), and the Paluch group (Katowice, Poland), part were detailed in previous publications involving some of the authors of this paper, part were measured for this paper at three different experimental setups in our labs at Roskilde and Tempe. The three setups used are briefly described below, where the new measurements are also presented. If nothing else is noted, chemicals were purchased from Sigma-Aldrich Chemical Company and used as acquired -- most of them are moderate-viscosity liquids at room temperature. 

{\it Roskilde University Setup}, (RU setup).
The dielectric cell is a multilayered gold-plated capacitor with empty capacitance $71$ pF. The capacitance was measured with an HP 3458A multimeter in the range of $10^{-3}-10^2$ Hz in conjunction with an HP 4284A LCR meter used in the frequency range $10^2-10^6$ Hz. The multimeter measurements were performed on a homebuilt setup that consists of a voltage divider involving the multimeter in combination with a homebuilt arbitrary wavefunction generator \cite{RosSetup2}. The latter produces low-frequency ($10^{-3}-10^2$ Hz ) sinusoidal signals with voltages that are reproducible within 10 ppm \cite{RosSetup2}.

The sample was placed into a homebuilt nitrogen-cooled cryostat which has absolute temperature accuracy better than 0.2 K and temperature stability during measurement better than 20 mK. The two measuring devices are connected to the measuring cell through a mechanical switch between the two frequency ranges (applied at 100 Hz). To ensure that the liquids were in thermal equilibrium after a temperature step, we waited 20 minutes before each measurement. Two frequency scans were taken at each temperature; data were only accepted if no differences were observed between the two spectra (beyond noise).

\begin{center}
\epsfig{file=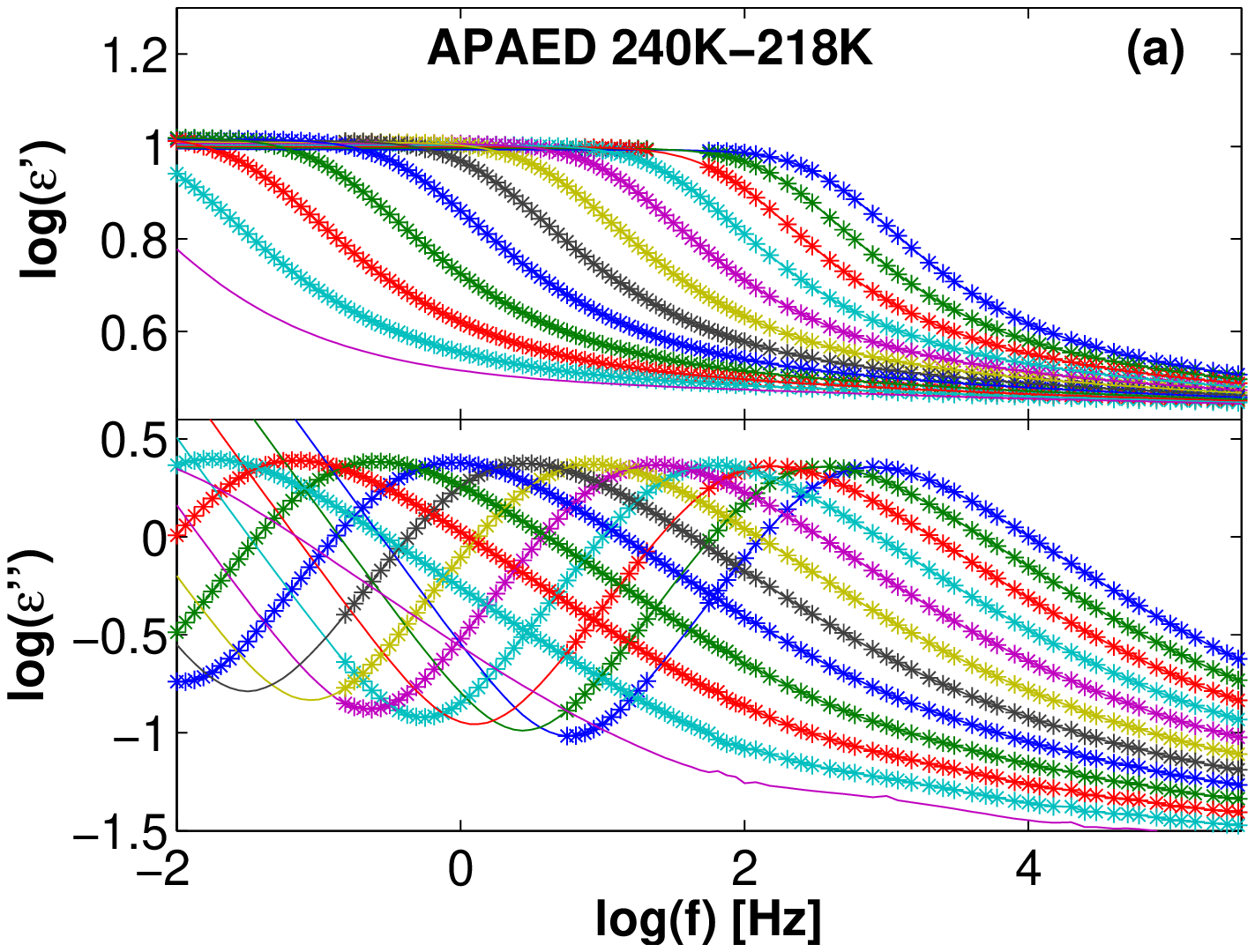,width=0.49\linewidth}
\epsfig{file=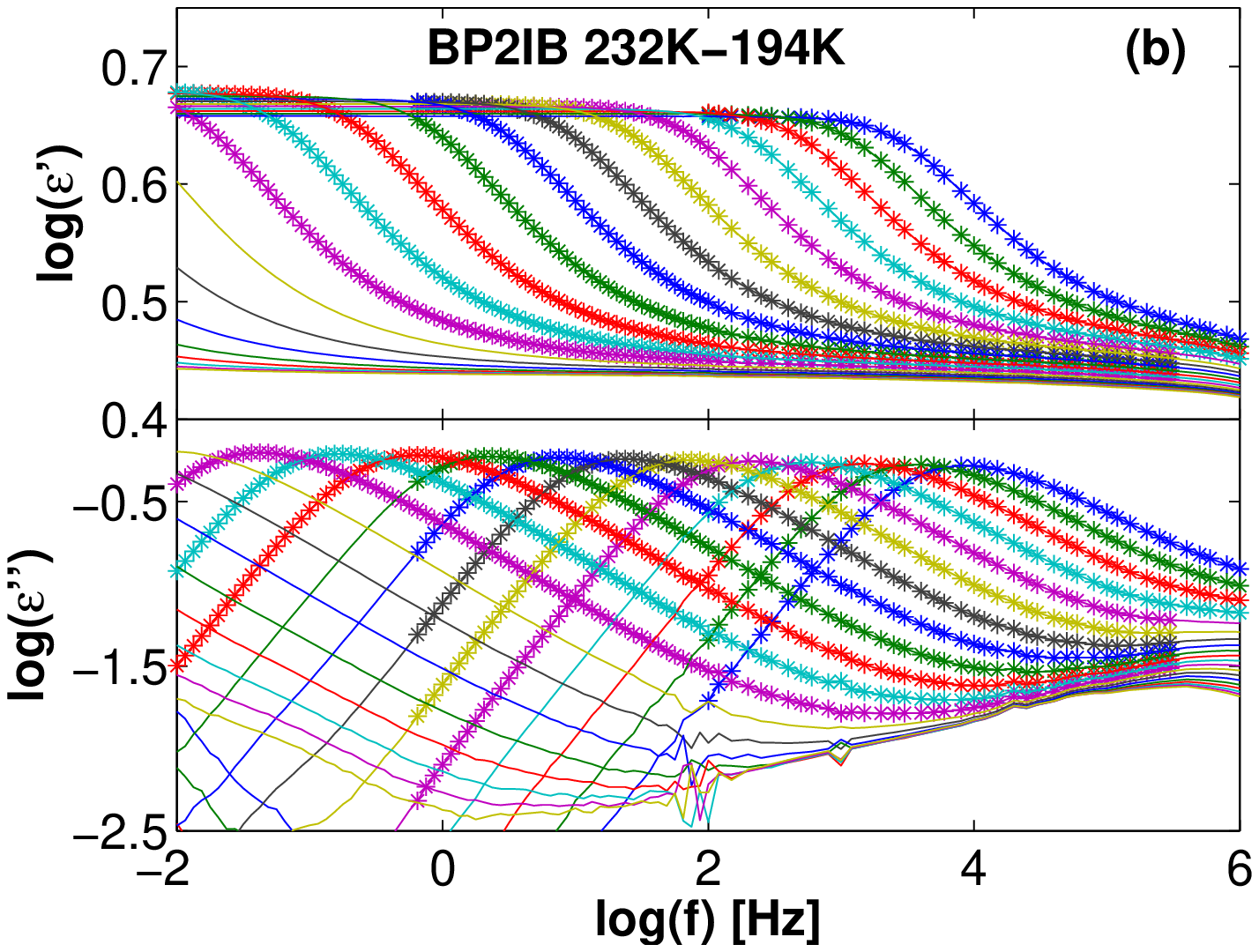,width=0.49\linewidth}
\epsfig{file=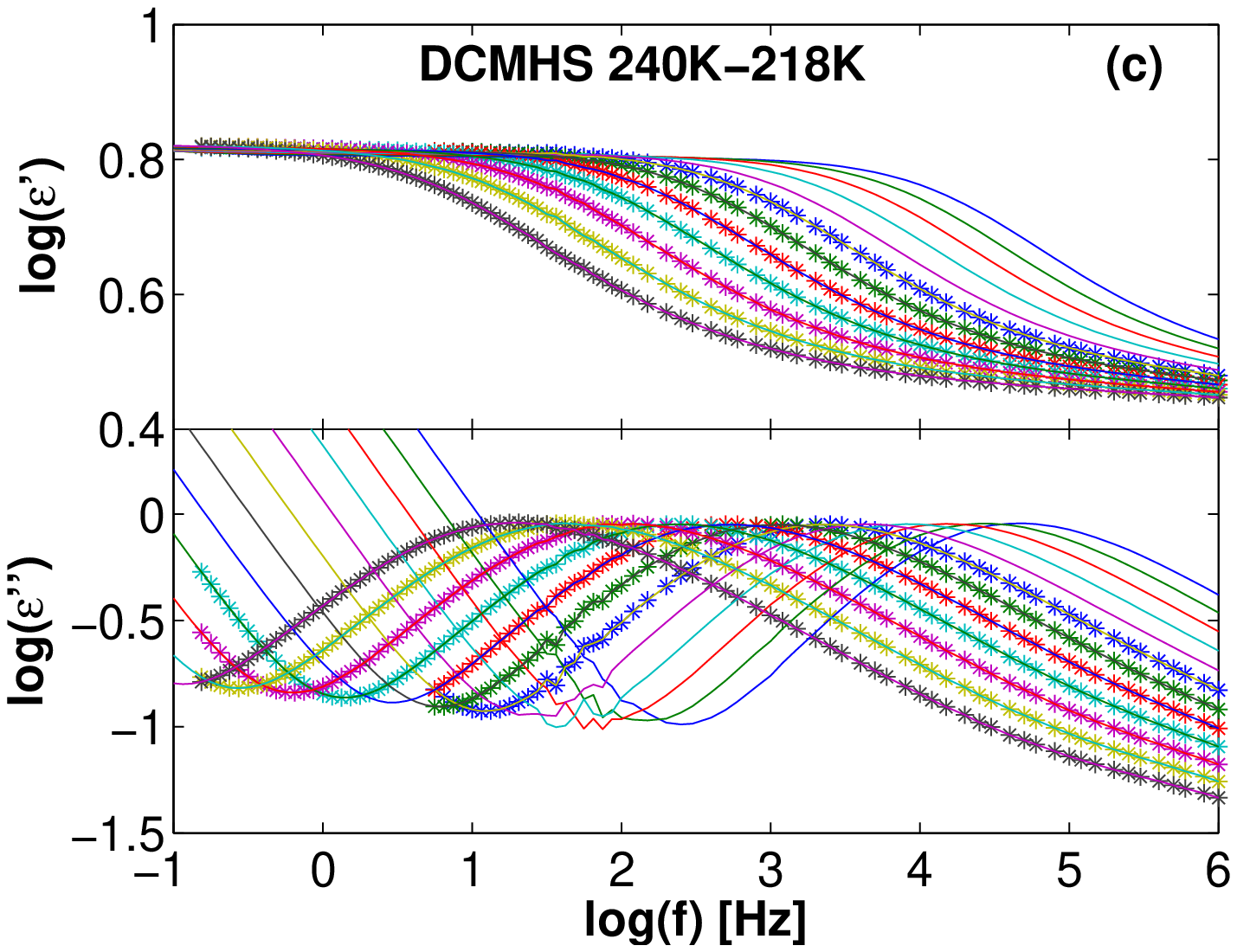,width=0.49\linewidth}
\epsfig{file=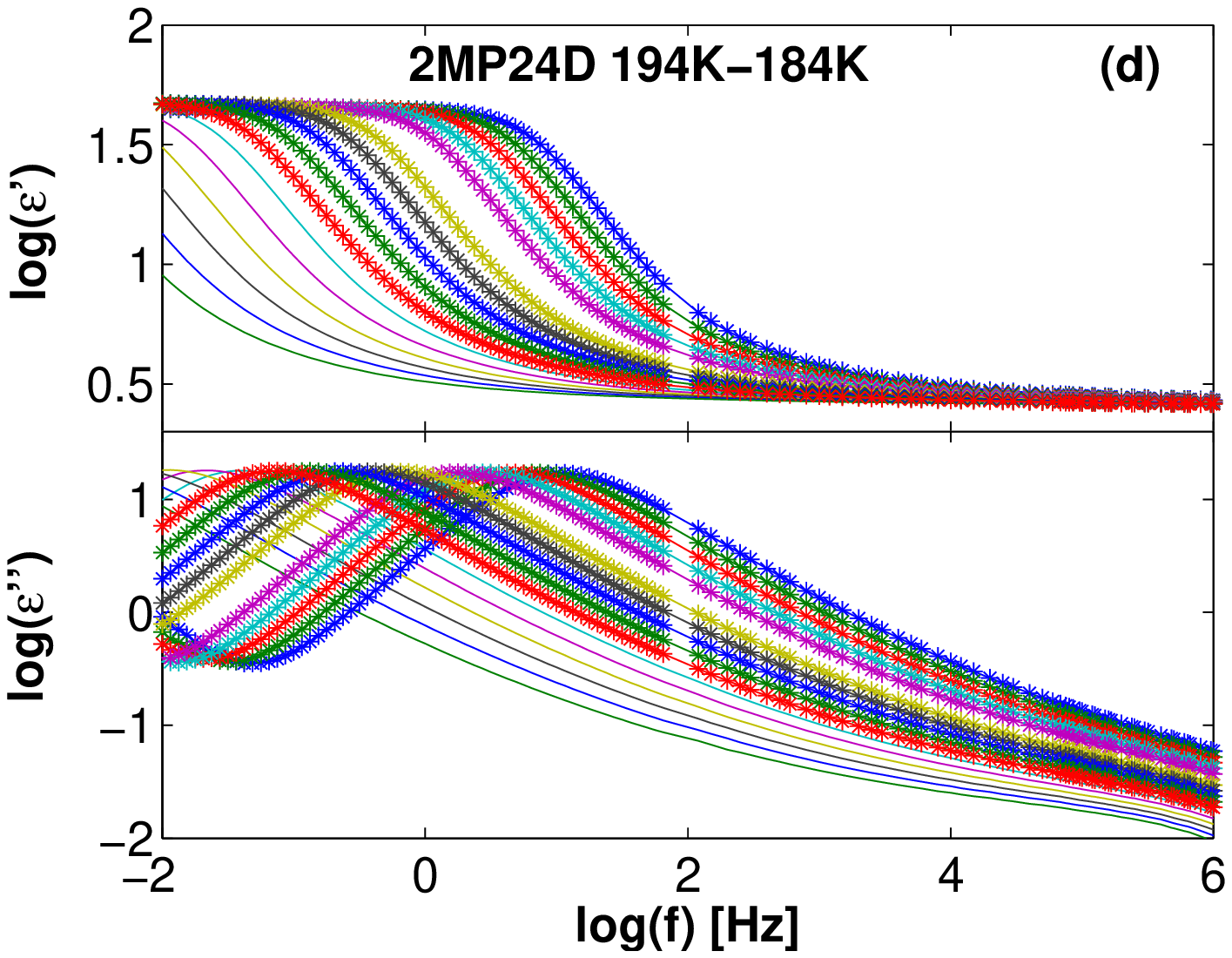,width=0.49\linewidth}
\epsfig{file=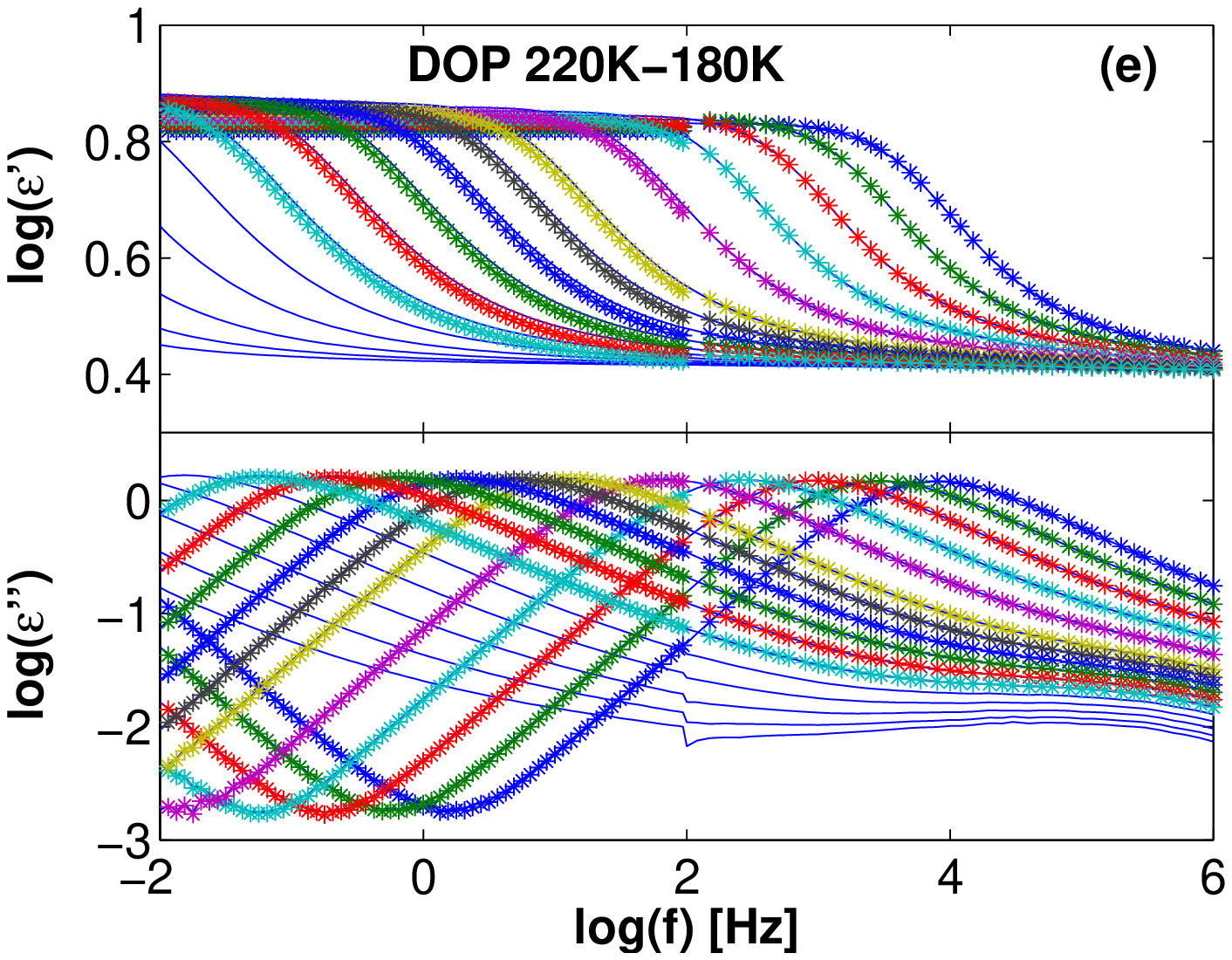,width=0.49\linewidth}
\epsfig{file=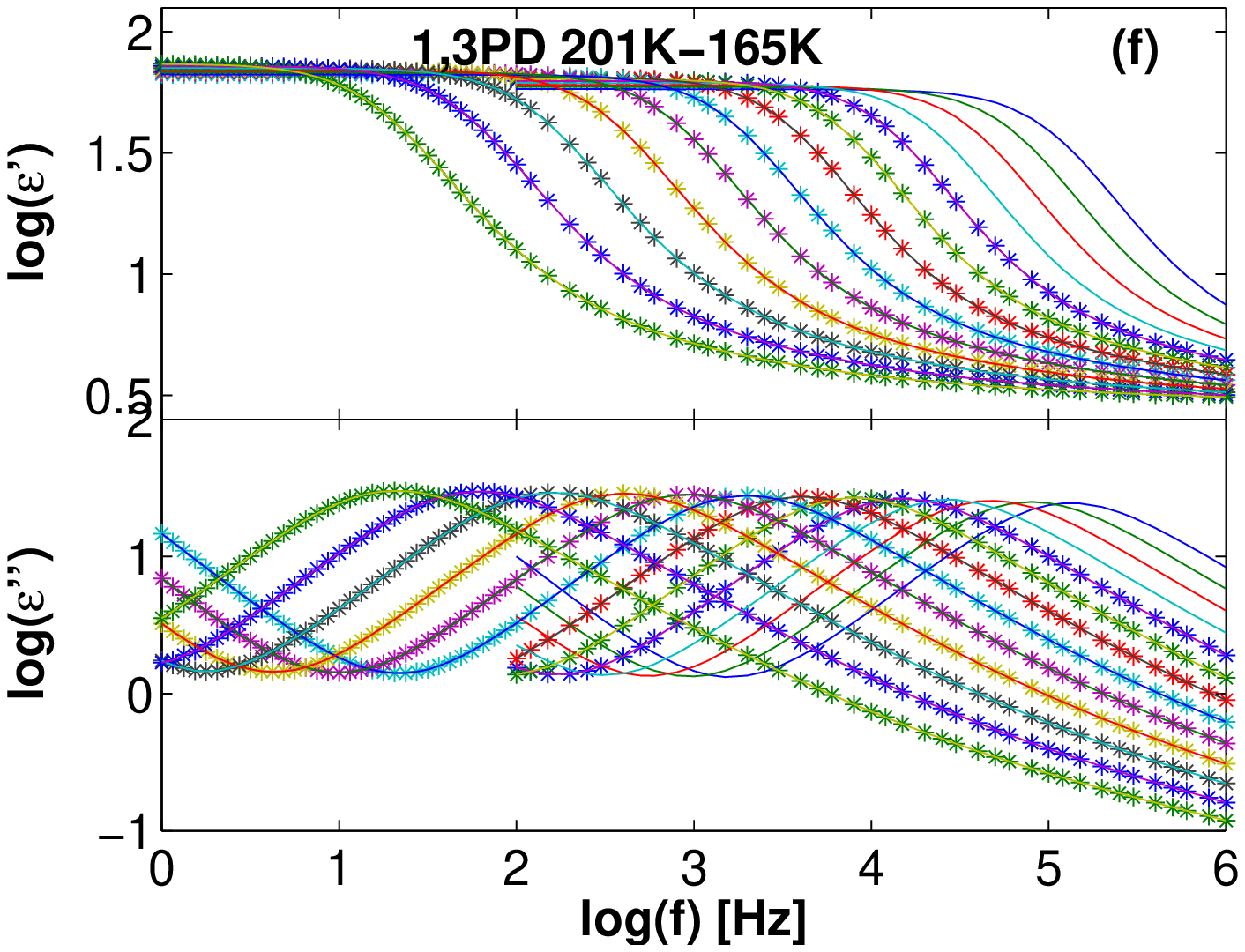,width=0.49\linewidth}
\epsfig{file=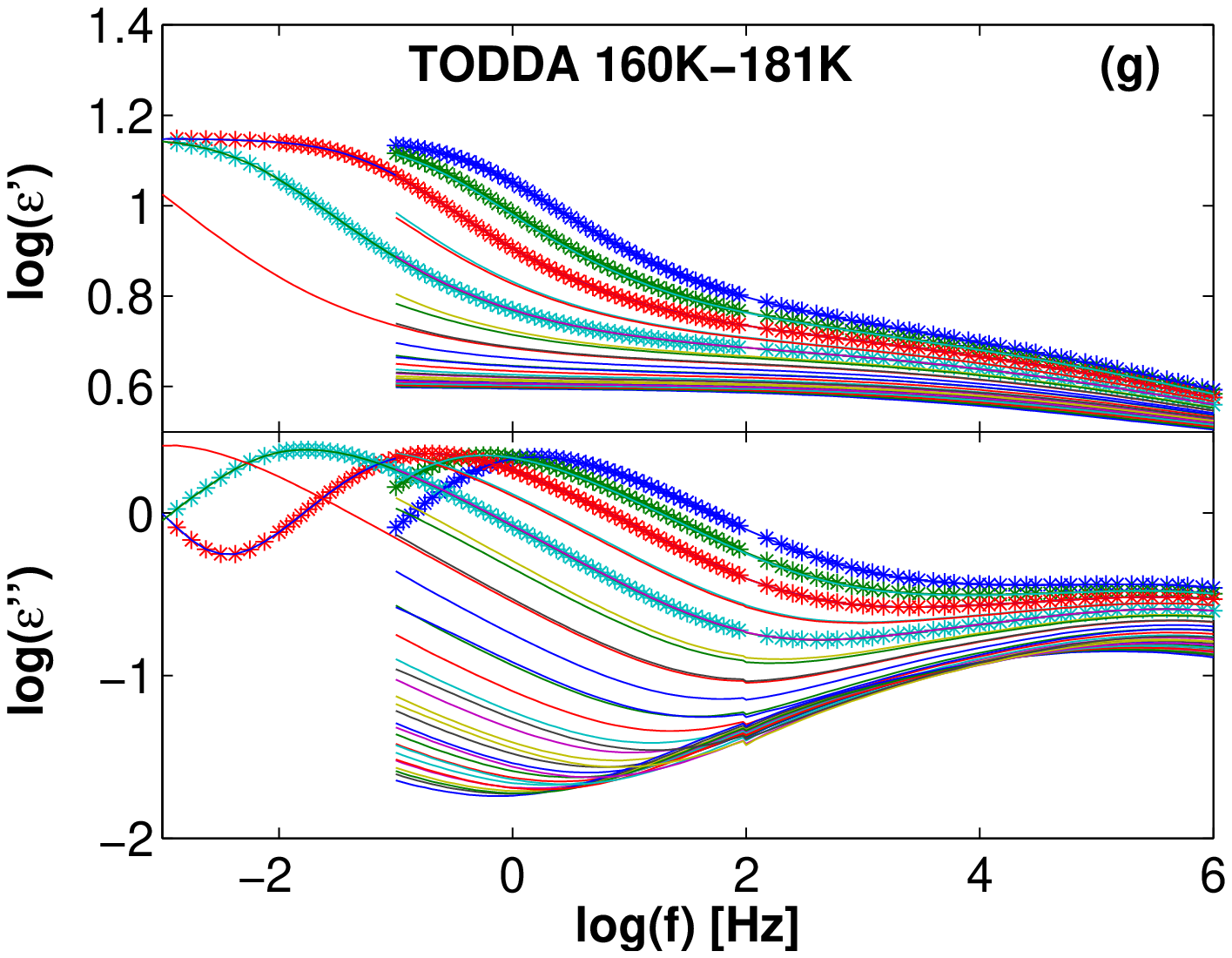,width=0.49\linewidth}
\epsfig{file=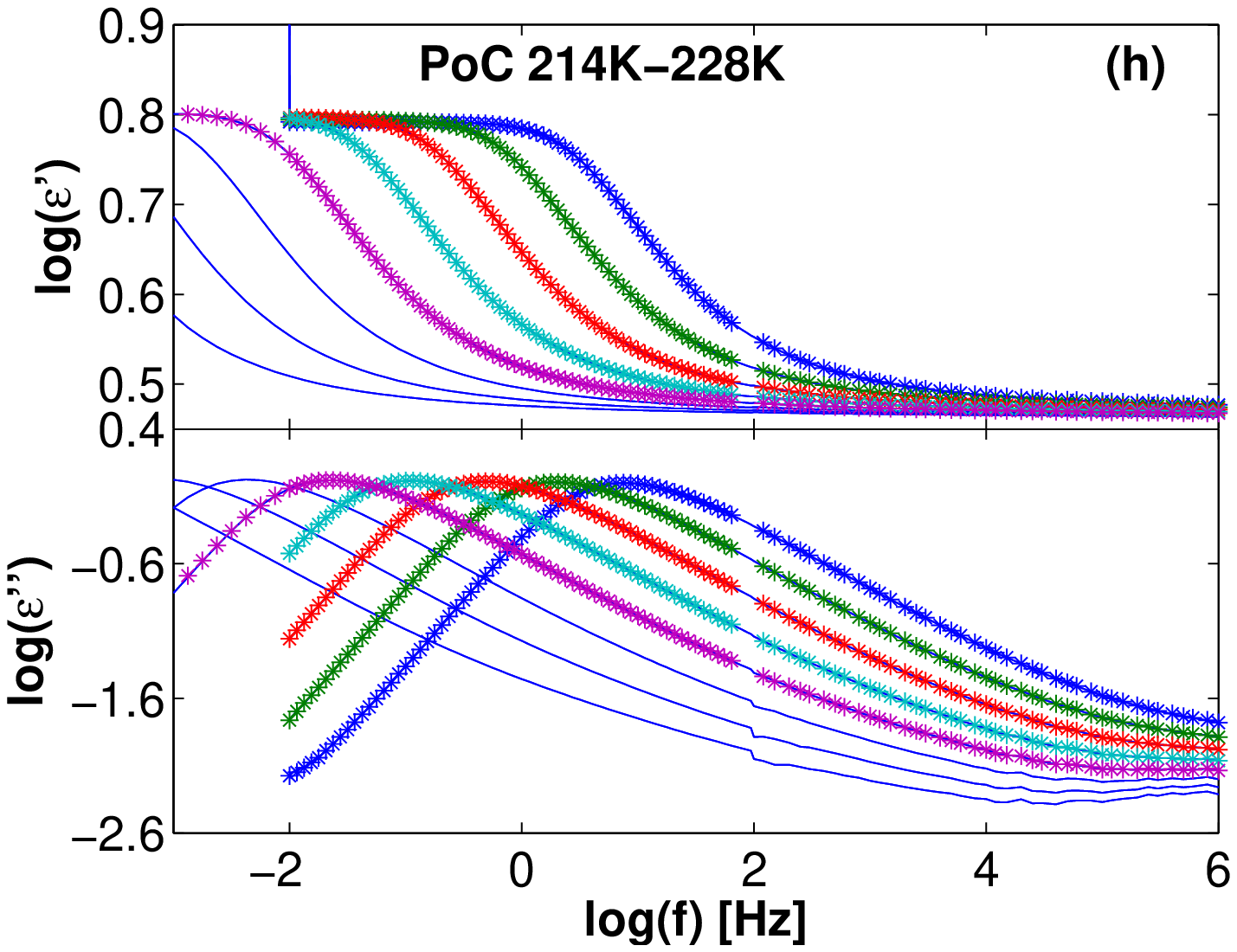,width=0.49\linewidth}
\epsfig{file=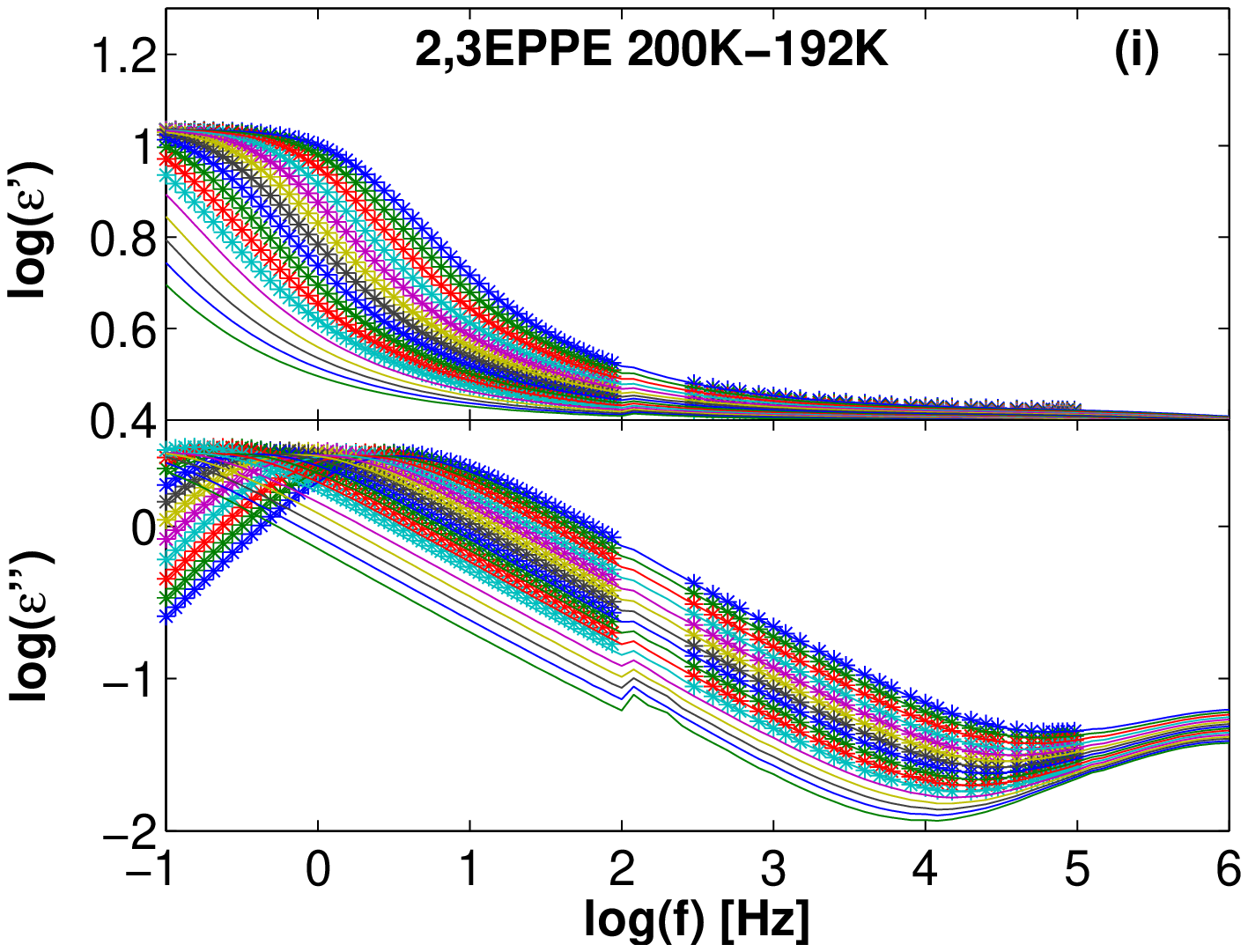,width=0.49\linewidth}
\epsfig{file=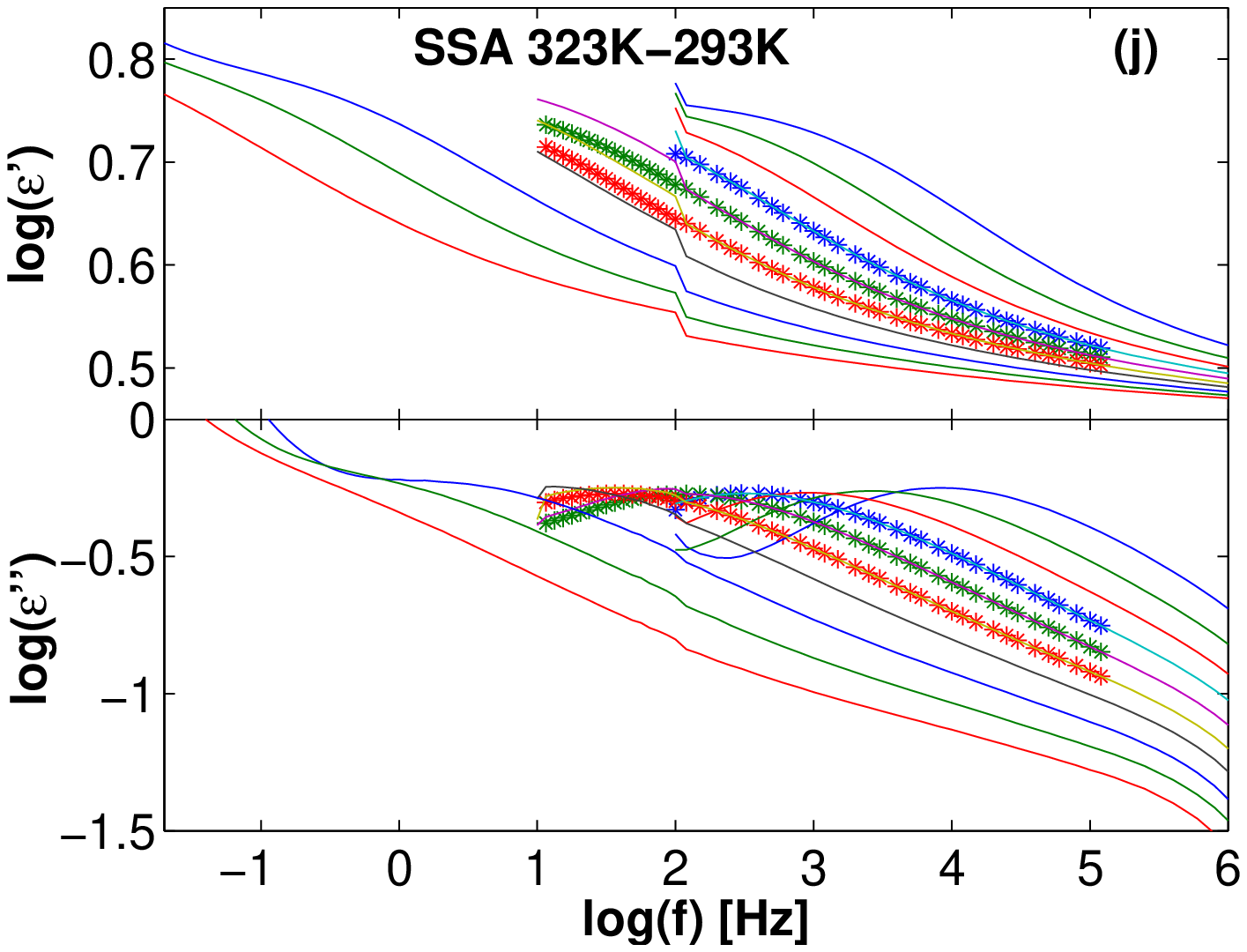,width=0.49\linewidth}
\epsfig{file=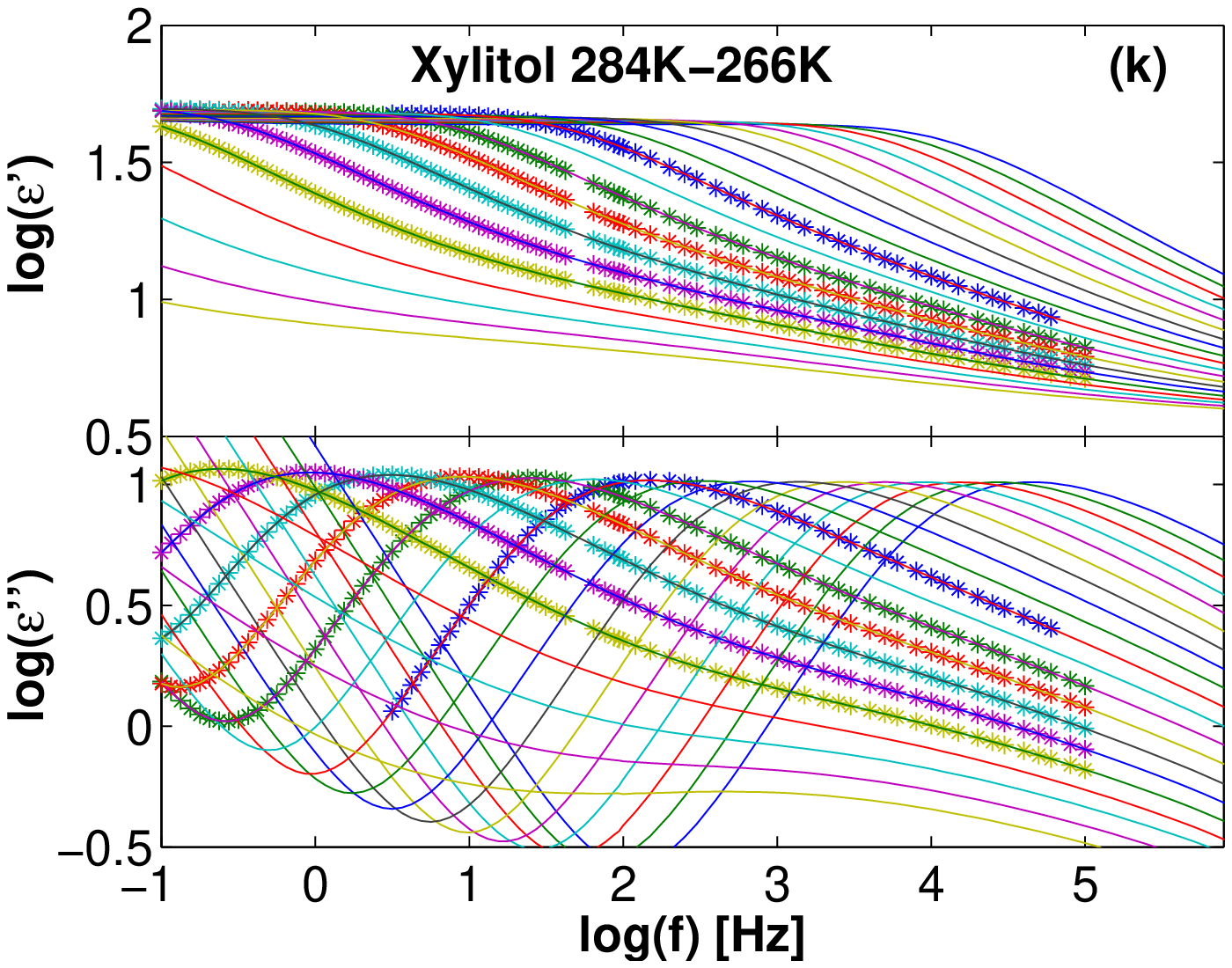,width=0.49\linewidth}
\epsfig{file=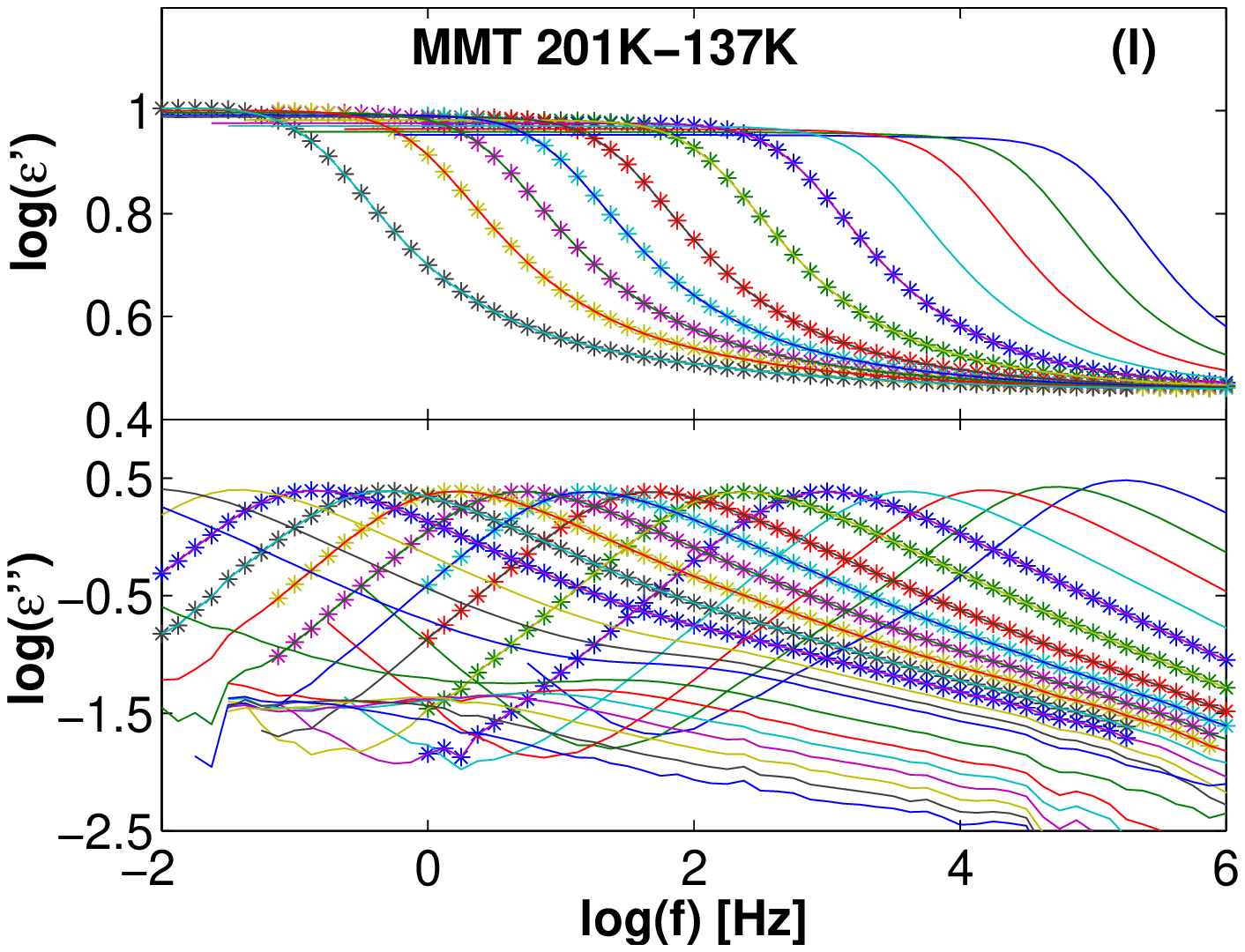,width=0.49\linewidth}
\epsfig{file=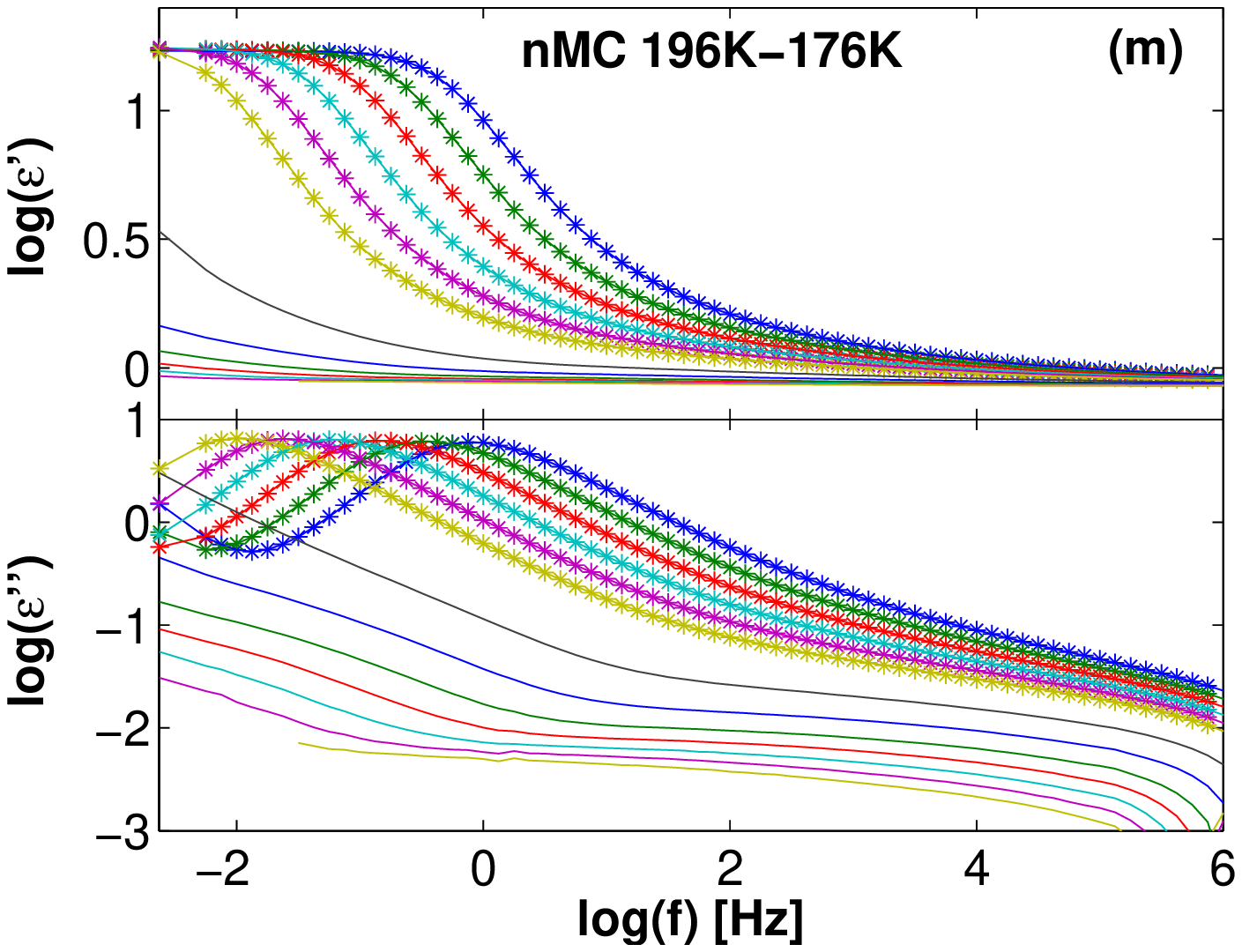,width=0.49\linewidth}
\epsfig{file=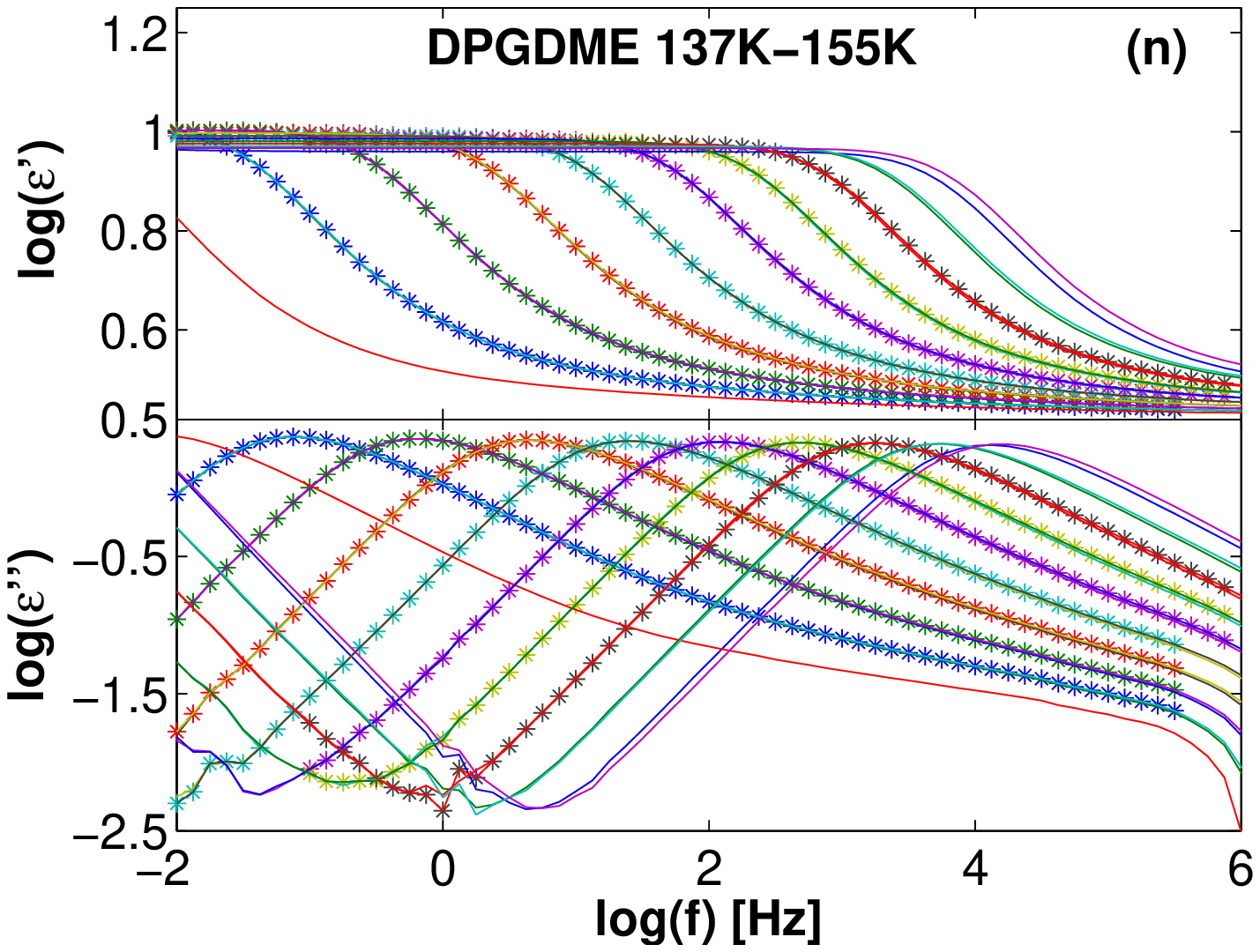,width=0.49\linewidth}

\end{center}
\pagebreak

\begin{figure}[t!]
\begin{center}
\epsfig{file=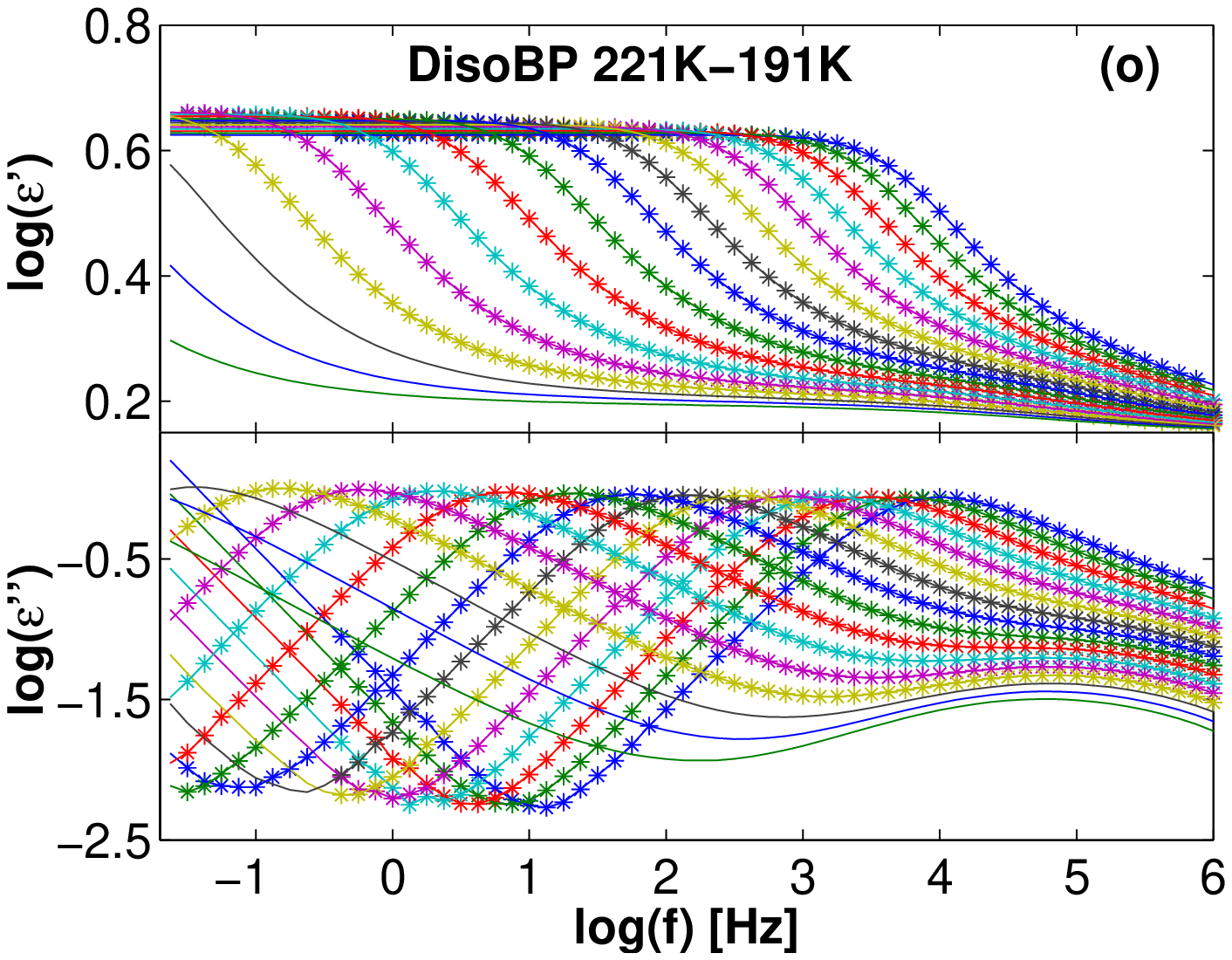,width=0.49\linewidth}
\epsfig{file=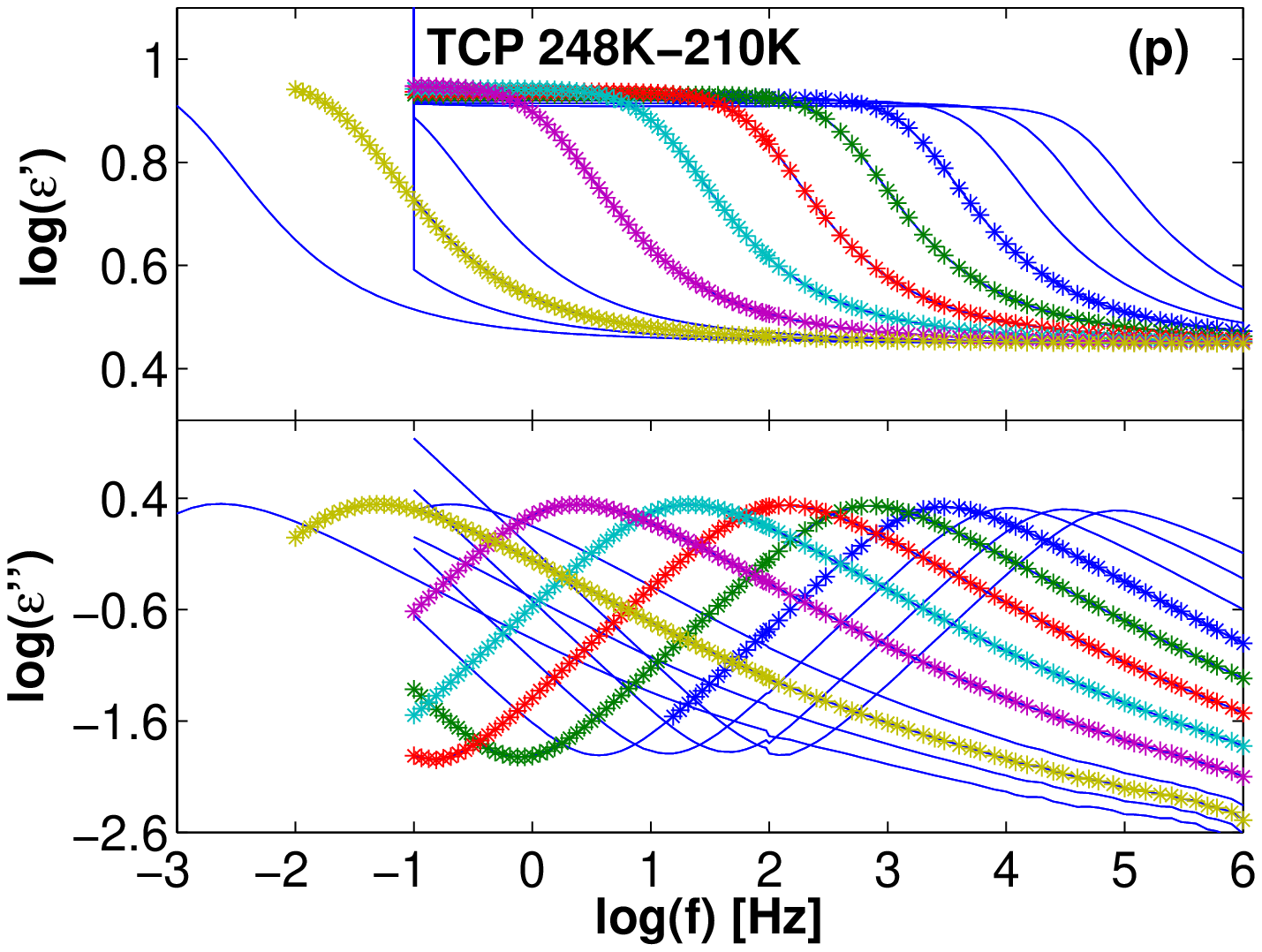,width=0.49\linewidth}
\epsfig{file=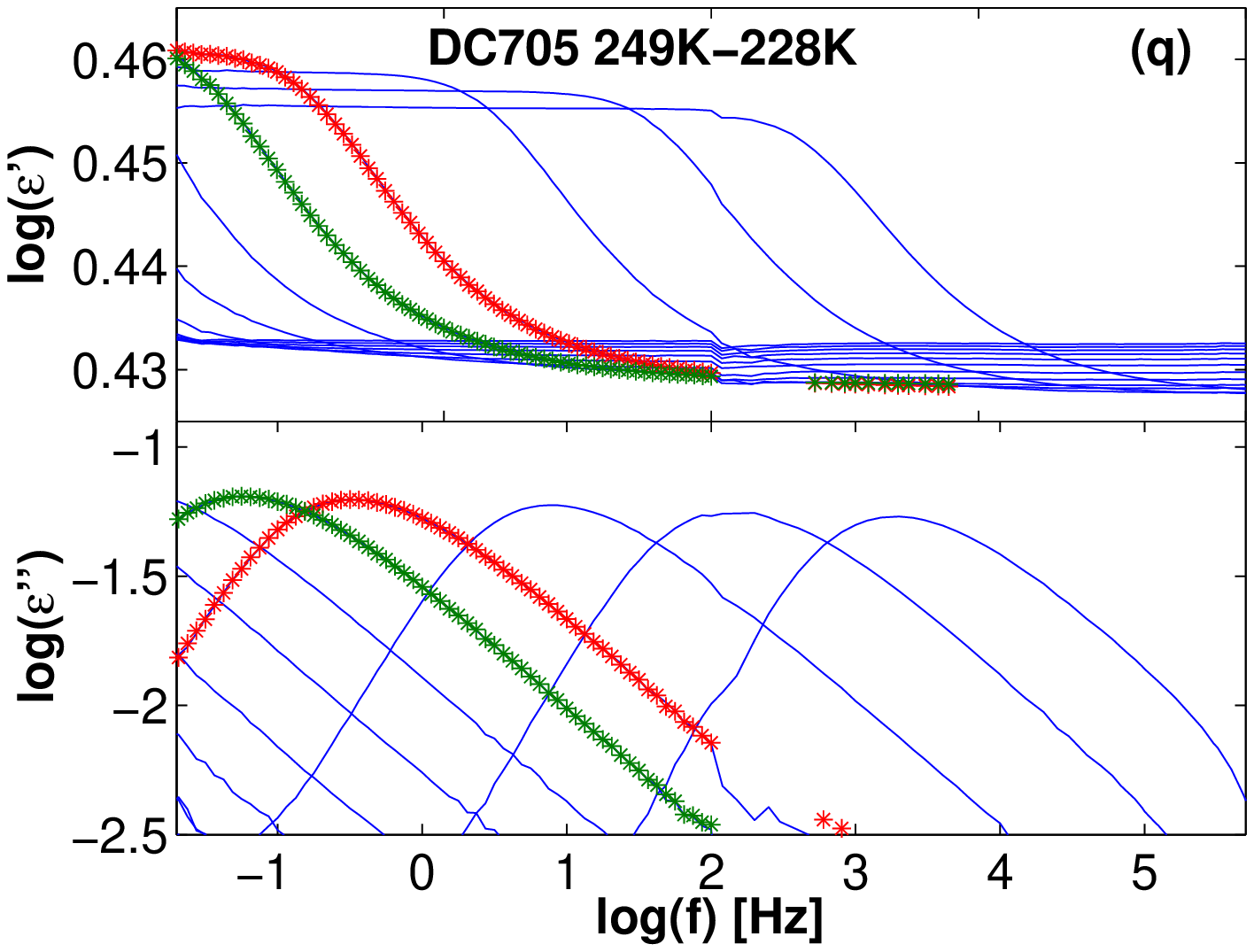,width=0.49\linewidth}
\end{center}
 \caption{\label{fig:raw_data1} Frequency-temperature scans for  
(a) 2-phenyl-5-(acetomethyl)-5-ethyl-1,3-dioxa-cyclohexane (APAED), 
(b) biphenyl-2-yl isobutylate (BP2IB)
(c) dicyclo-hexyl-2-methyl succinate (DCHMS), 
(d) 2-methyl-pentane-2,4-diol (2MP24D), 
(e) dioctyl phthalate (DOP), 
(f) 1,3 propandiol (13PD), 
(g) trioxatridecane diamine (TODDA), 
(h) $\alpha$ phenyl-\textit{o}-cresol (PoC), %, (g)silicone oil MS407, 
(i) 2,3-epoxypropyl phenylether (2,3EPPE), 
(j) salicylsalicylic acid (SSA) and 
(k) xylitol.
(l) methyl-\textit{m}-toluate (MMT). 
(m) \textit{N}-$\epsilon$-methyl-caprolactam  (nMC) 
(n) dipropylene glycol dimethyl ether (DPGDME) 
(o) di-\emph{iso}-butyl phthalate (DisoBP) 
(p) tricresyl phosphate (TCP) and 
(q) trimethyl-pentaphenyl trisiloxane (DC705).
The full curves give the temperature-frequency scans, stars mark the data and corresponding data range selected for the analysis. On the plots (a)-(k) there is a systematic error around $100 $ Hz due to the supply net frequency and the fact that we at $100 $ Hz switch between two measuring techniques.\label{data}}
\end{figure}

The following liquids (with noted purity, abbreviation, and figure) were measured on this setup:  
2-phenyl-5-(acetomethyl)-5-ethyl-1,3-dioxacyclohexane (APAED, Fig. \ref{fig:raw_data1}(a)), 
biphenyl-2-yl isobutylate (BP2IB, Fig. \ref{fig:raw_data1}((b)), 
dicyclo-hexyl-2-methyl succinate (DCHMS, Fig. \ref{fig:raw_data1}(c)), 
(the three liquids (a), (b), and (c) were synthesized at D\'{\i}az-Calleja's laboratory at Universidad Politécnica de Valencia); 
2-methyl-pentane-2,4-diol ($98\%$, British Drug Houses Ltd., 2MP24D, Fig. \ref{fig:raw_data1}(d)),
dioctyl phthalate ($99\%$, DOP,  Fig. \ref{fig:raw_data1}(e), 
1,3 propandiol ($98\%$, 13PD,  Fig. \ref{fig:raw_data1}(f)),
trioxatridecane diamine (TODDA,  Fig. \ref{fig:raw_data1}(g)), 
$\alpha$ phenyl-\textit{o}-cresol ($98\%$, PoC,  Fig. \ref{fig:raw_data1}(h)), 
2,3-epoxypropyl phenylether ($99\%$, 2,3EPPE,  Fig. \ref{fig:raw_data1}(i)), 
tricresyl phosphate ($98\%$, Alfa Aesar, TCP, Fig. \ref{fig:raw_data1}(p), data for structural relaxation times published in \cite{Hecksher08}), 
trimethyl-pentaphenyl trisiloxane (Dow Corning 705 silicon diffusion pump oil, Dow Corning Corp., DC705, Fig.  \ref{fig:raw_data1}(q)), 
1,2 propandiol ($99\%$, Merk, PG ),
dibutyl phthalate ($98\%$, DBP), and 
diethyl phthalate ($97\%$, DEP). 
-- Spectra for these liquids are shown in Fig. \ref{fig:raw_data1} except for the last three liquids that have often been reported in the literature.
Salicylsalicylic acid ($99 \%$, SSA, Fig. \ref{fig:raw_data1}(j)), xylitol ($\geq 99 \%$, Fig. \ref{fig:raw_data1}(k)) and D(-)sorbitol ($99 \%$, AppliChem, Sor) are crystals at room temperature. They were melted in an oven, placed in the warmed-up (melting temperature) capacitor and subsequently cooled to room temperature. Xylitol was kept at $370$ K for one hour; D(-)sorbitol at $390$ K for four hours; SSA kept at $419$ K for one hour. All other liquids were cooled starting from room temperature.

{\it Arizona State University Setup 1}, (ASU Setup1).
This setup is basically described in Refs. \cite{Richert1995, hansen1997, Huang2005}, but used here with some recent improvements. The measuring cell has empty capacitance $17$ pF. The sample cell was placed on a temperature-controlled plate in an evacuated He-refrigerator cryostat (Leybold RDK 6-320) driven by a Cool Pak 6200 compressor. The temperature of the base plate and the cell was controlled by a Lakeshore 340 temperature controller equipped with calibrated DT-470-CU diodes as sensors. The capacitance cell was connected to a Solartron SI-1260 gain/phase analyzer equipped with a Mestec DM-1360 trans-impedance amplifier \cite{Richert1995}.  The liquids were supercooled in the cryostat chamber. Due to the relatively low cooling rate, around $1.5 \, \textmd{K/min}$,  the waiting time  between a temperature step and the start of measurements was $10$ minutes after $5$ minutes temperature stabilization.

The following liquids (characterized by particularly low glass transition temperatures) were measured on this setup:  
2-methyltetrahydrofuran ($99.1\%$, \emph{distilled}, MTHF), 
methyl-\textit{m}-toluate ($98\%$, Avocado Research Chemicals Ltd., MMT, Fig. \ref{fig:raw_data1} (l)) and
n-propyl-benzene ($99\%$, nPB).

{\it Arizona State University Setup 2}, (ASU Setup2).
The measuring cell, which has empty-cell capacitance $27$ pF, consists of two steel discs electrodes of diameter $20$ mm separated by six $50 \mu$m thick Teflon stripes. The cell was placed inside a nitrogen-gas cooled cryostat where temperature was stabilized and measured by a Novocontrol Quatro controller. The impedance measurements were performed in the range $0.1$ Hz - $10$ MHz using a Solartron SI-1260 gain-phase analyzer. A Mestec DM-1360 trans-impedance amplifier was used (as for ASU Setup 1). The empty sample capacitor
was used as reference to calibrate the frequency-dependent trans-impedance of the amplifier.

The following liquids were measured on this setup: 
\textit{N}-$\epsilon$-methyl-caprolactam  ($99\%$, nMC, Fig. \ref{fig:raw_data1} (m)), 
dipropylene glycol dimethyl ether ($\geq 98\%$, DPGDME, Fig.\ref{fig:raw_data1} (n)) 
and  di-\emph{iso}-butyl phthalate ($99\%$, DisoBP, Fig. \ref{fig:raw_data1} (o)).

The data to our disposition were thus obtained on several different setups working in different frequency intervals with varying number of measurement frequencies per decade. From the spectra measured at the RU setup we removed the points around $100$ Hz because of the systematic error due to the switch; all other data sets were used as measured, or received from the different groups. If two data series for the same liquid were available from different groups/setups, the series with most frequencies measured per decade was used. 

Decahydroisoquiline (DHIQ) is represented by two datasets, one measured by Jakobsen et al \cite{Jacobsen05} (RU Setup) and one by Richert et al \cite{richert2003} (ASU Setup2). These measurements compliment each other nicely, except for a minor deviation ($\sim0.5$ K) in the absolute temperature calibration.

Following a basic philosophy of analyzing the raw data directly, no attempts were made to subtract contributions from the DC conductivity and no attempts were made to subtract contributions from $\beta$ relaxation(s). This procedure is fundamental to this paper's approach. Thus while one may argue what is the correct way of compensating for these and possibly other interfering effects in order to isolate the ``true'' $\alpha$ process, it should be much easier to reach consensus regarding the raw data themselves and their properties.

\section{Data analysis methods}

\begin{figure}[tb!]
\begin{center}
\epsfig{file=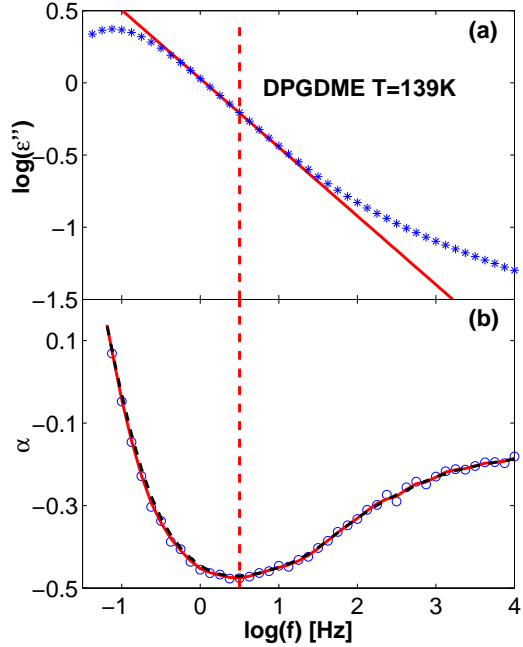,width=0.45\linewidth}
\end{center}
 \caption{Illustration of the procedure used to calculate the minimum slope.
(a) Data for the dielectric loss, $\eps ''$, of dipropylene glycol dimethyl ether (DPGDME) at $T=139$ K in double-logarithmic plot ({\color{MycBlue} $*$}). The red line marks the inflection point tangent that has slope equal to the minimum slope $\am$. (b) The calculated values of the slope by numerical differentiation from these data ({\color{MycBlue} $\circ$}). The red curve marks the slope data after averaging twice (over two neighboring points), the dashed line is after ten applications of the averaging routine. The vertical dashed line through both plots marks the position of the minimum slope frequency. \label{fig:alpha:ana}}
\end{figure}

The minimum slope of the dielectric loss plotted in a log-log plot is identified directly from raw data; thus no assumptions concerning the nature of the relaxation process are made, for instance of how  $\alpha$ and $\beta$ processes interact, whether or not the excess wing is a hidden $\beta$ process, etc. \cite{gai2006}. The slope in the log-log plot is given by 

\begin{equation}
 \alpha=\frac{d \log \eps''}{d \log (f) }\,,
\end{equation}
where $f$ is the frequency. Figure \ref{fig:alpha:ana} illustrates the minimum slope concept by showing the high-frequency imaginary part of the complex dielectric constant (upper panel) at a given temperature for one liquid (DPGDME, $T=139$ K) and, in (b), the corresponding slope where, of course, $\alpha=0$ at the loss peak frequency $\fm$. The minimum of the derivative above the loss peak frequency defines the minimum slope, $\am$, which is always a negative number.

Since the second-order derivative is by definition zero where the slope is minimal, at  the inflection point,  the linear tangent approximation works particularly well here. This means that the approximate power-law description $\eps''\propto f^{\am}$ gives a good representation of the high-frequency loss over a sizable frequency range. Thus if, for instance, the minimum slope $\am$ is close to $-1/2$, then to a good approximation $\eps''\propto 1/\sqrt{f}$ for $f>>\fm$ over a significant frequency range. In the time domain this corresponds to $\sqrt t$ relaxation being a good approximation of the relaxation function. 

To determine the minimum slope for a given data set, the set was first numerically point-by-point differentiated. Only data sets with a well-defined minimum slope -- or a clear plateau of constant slope -- were included in the analysis. Moreover, data sets were only included if there was so little noise in the resulting slopes that determination of $\am$ with two significant digits was possible. These selection criteria imply that several frequency scans at high temperatures, as well numerous noisy data sets, were eventually omitted from the data analysis. 

As a means to increase the reliability of the $\am$ estimate we applied averaging. Thus the noise in the numerical derivative was reduced by repeatedly applying a routine that averages over two neighboring points. The number of times this averaging procedure was applied varied with the data set, but kept below ten. As an example, for the data in Fig. \ref{fig:alpha:ana}(b) a double iteration of the averaging routine was used; the black dashed line shows the result if averaging was instead applied ten times. If averraging ten times changed $\am$ more than $0.01$, the data set was discarded. Subsequent applications of the smoothing procedure result in numerically slightly larger values of the minimum slope, but this was never a serious problem. If the resulting curve after ten averagings was still too noisy, the frequency scan was discarded. Thus some subjectivity enters the analysis, but we took care to keep the element of subjectivity as small as possible; whenever questions arouse making the applied procedure dubious, the data set was discarded. This procedure left a total of $53$ liquids in the data collection out of an initial collection of $84$ liquids; for each liquid the number of identified minimum slope values varies between $2$ and $17$ with values ranging from $-0.75$ up to $-0.10$. Altogether $347$ minimum slopes were identified for the $53$ liquids at various temperatures.

The above-described approach for the characterization of the high-frequency relaxation was adopted in order to be as objective as possible by avoiding the need to make a choice of fitting intervals. In latter case it is necessary if one fits data to, e.g., a stretched exponential (KWW) or CD function to decide below which frequency the $\alpha$ process is no more likely to be affected by secondary ($\beta$) processes. The subjectivity in the choice of fitting intervals results in  numbers $\beta_{KWW}$ and $\beta_{CD}$ that are in many cases not uniquely determined with two-decimal accuracy -- giving the same data set to different people will generally result in slightly differing fitting parameters.

\begin{figure}[tb!]
\epsfig{file=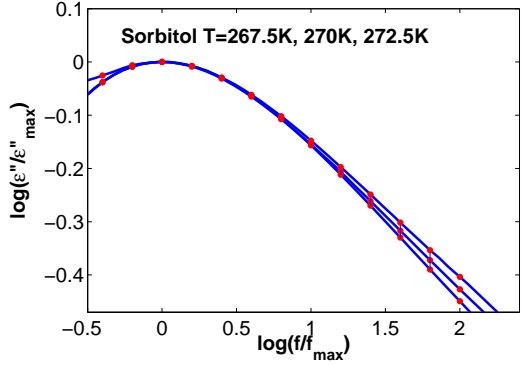,width=0.46\linewidth}
  \caption{Master plot of sorbitol data taken at $267.5$, $270$ and $272.5$ K, i.e., log-log plot of the data normalized to have maximum equal to unity at unity normalized frequency. The TTS measure $\Delta$ is defined as the area difference between two neighboring temperature curves in this plot (where the area is calculated by including 0.4 decade of lower and two decades of higher frequencies than the loss peak frequency)  divided by the difference of the logarithms of the actual loss peak frequencies (Eq. (\ref{delta_def})). The points on the graphs mark the $\varepsilon '' $ values used for the calculation of $\Delta$
\label{fig:norm}}
\end{figure}

We need one further parameter to characterize the shape of the loss peak. For this we choose the width at half loss measured in decades. In order to be able to make optimal use of the data sets (that are often significantly affected by the existence of the DC contribution to the left of loss peak) we used only the number of decades of frequency to the right of the loss peak frequency until the loss is halved. The obtained widths are conveniently normalized with respect to the half Debye width on the log scale, $W_{D/2}= 0.571$. Thus if the observed half width on the log scale is $W_{1 /2}$, we define

\begin{equation}\label{wdef}
\wh\,\equiv\, 
\frac{W_{1/2}}{W_{D/2}}\,.
\end{equation}
This quantity is always above unity; if it is close to one, the relaxation is Debye like.

Turning to the quantification of how well time-temperature superposition (TTS) applies, we note that to decide whether TTS applies one usually uses a visual evaluation of attempted master plots of losses measured at different temperatures. One way to evaluate TTS is to investigate whether shape parameters are temperature invariant; however as mentioned we wish to avoid the use of fits to analytical functions. In order to obtain a numerical measure of how well TTS applies, the width variation with temperature is first quantified as follows. Consider loss spectra at two neighboring temperatures, $T_j < T_{j+1}$, both normalized with respect to their respective loss peak frequencies $\fm$ and amplitude $\eps''_{max}$ (identified by fitting a second-order polynomial to an interval of data points in double logarithmic plot, using from 5 up to 9 points around the maximum depending on the symmetry of the loss peak). The difference between the two normalized curves is reflected in the areas between the curves (Fig. \ref{fig:norm}). Let $ \tilde{\eps}=\eps''/\eps''_{max}$ and $\tilde{f}=f/f_{max}$ be normalized loss and frequency, respectively at a given temperature. We define $d S_j$ as the area between two frequency scans at $T_j $ and $T_{j+1}$: $d S_j$ is sum of the difference in the values of $\log(\tilde{\eps}_j)$ and $\log(\tilde{\eps}_{j+1})$ at $m$ frequencies in the normalized graphs. More precisely, we found $\eps''_j$ by interpolation at $m=13$ frequencies equally spaced on the logarithmic axis ranging from $\log(\tilde{f}_1)=-0.4$  to $\log(\tilde{f}_{13})=2.0$. The calculation of $d S_j$ and it was made with those 13  $\tilde{\eps}$ values,

\begin{equation}
d S_j= \sum_{i=1}^{13} \left| \log(\tilde{\eps}_{j+1}(\log(\tilde{f}_i)))-\log(\tilde{\eps}_{j}(\log(\tilde{f}_i)))\right|\,.
\end{equation}
To make reasonable sense the frequency interval $[\tilde{f}_1;\tilde{f}_{13}]$ should contain the main part of the $\alpha$ loss peak. We define this as including almost a half decade on the low-frequency side and two decades on the high-frequency side of the loss peak. The frequency-range asymmetry is justified by: 1) A wish to include as many dielectric spectra as possible at relatively low temperatures (i.e., in the low-frequency part of the experimental window) because many dissipation curves ends around 10mHz; the low temperature relaxation response is particularly interesting due to the separation of $\alpha$ and high-frequency $\beta$ processes; 2) An asymmetric interval reduces the effect of the DC contribution. -- Note that we need at least two frequency scans to calculate one value of the area difference and thus $\Delta$; thus the TTS analysis does not result in $347$, but in $347-53 =294$ data points.

The required measure of TTS deviations should not depend on the difference between neighboring temperatures in the particular data series under scrutiny. Thus, we define the TTS deviation measure $\Delta_j$ as follows (where $d \log (f_{max,j})$ is the numerical change in log(loss peak frequency))

\begin{equation}\label{delta_def}
\Delta_j\,=\, \frac{d S_j}{d \log (f_{max,j})}\,.
\end{equation}
In this way one compensates for the fact that measurements at close temperatures trivially result in curves of closely similar shapes. 

TTS is better obeyed, the smaller $\Delta$ is. This TTS measure introduces a further constraint on the data selection, namely that only data sets with a well-defined maximum and at least half a decade of measurements on the low-frequency side of loss were included in the analysis. Furthermore, data must be quite accurate since the $\tilde{\eps}$ values are found from data by linear extrapolation.

\section{Minimum slope distribution}

\begin{figure}[b!]
\begin{center} 
\epsfig{file=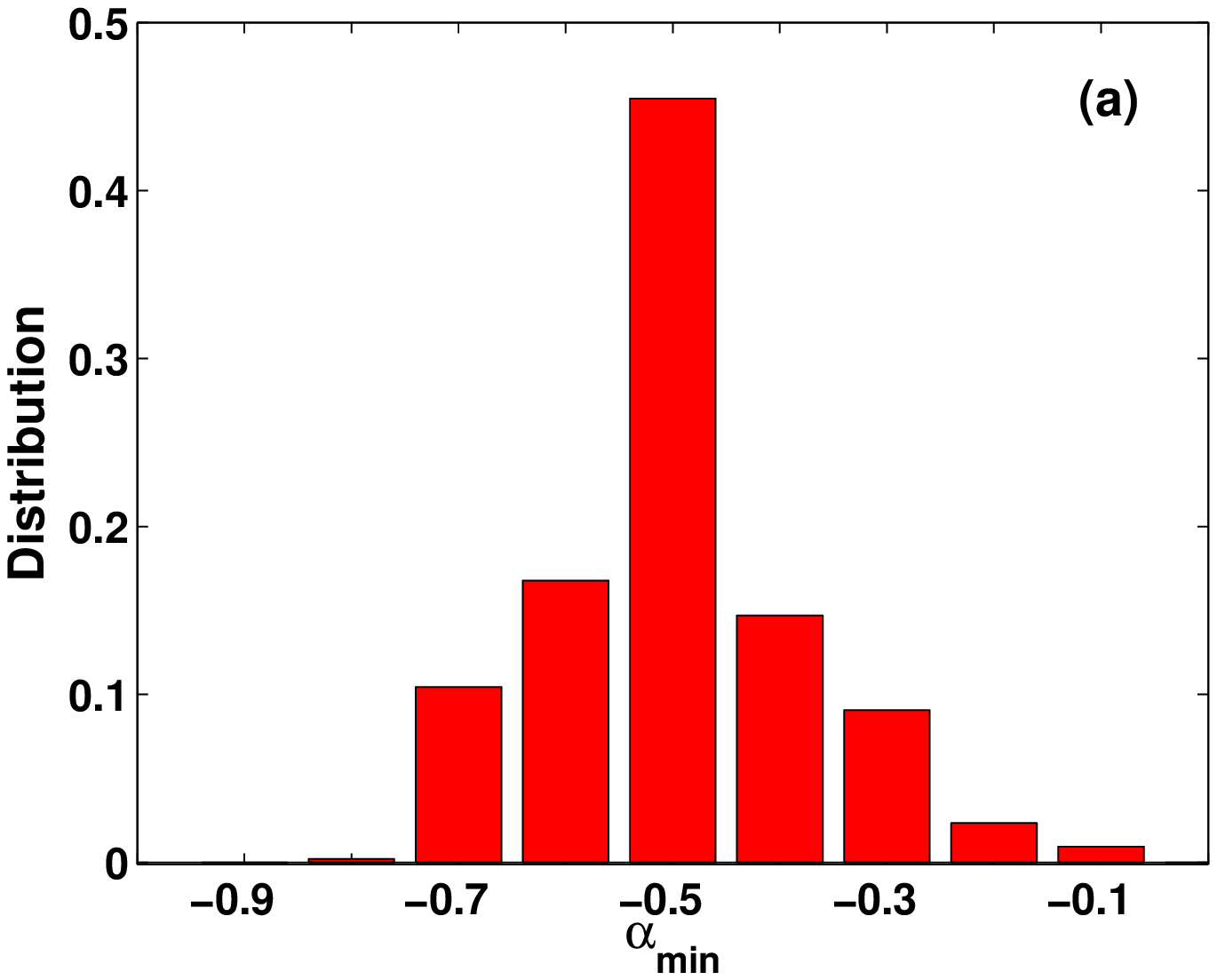,width=0.46\linewidth}
\epsfig{file=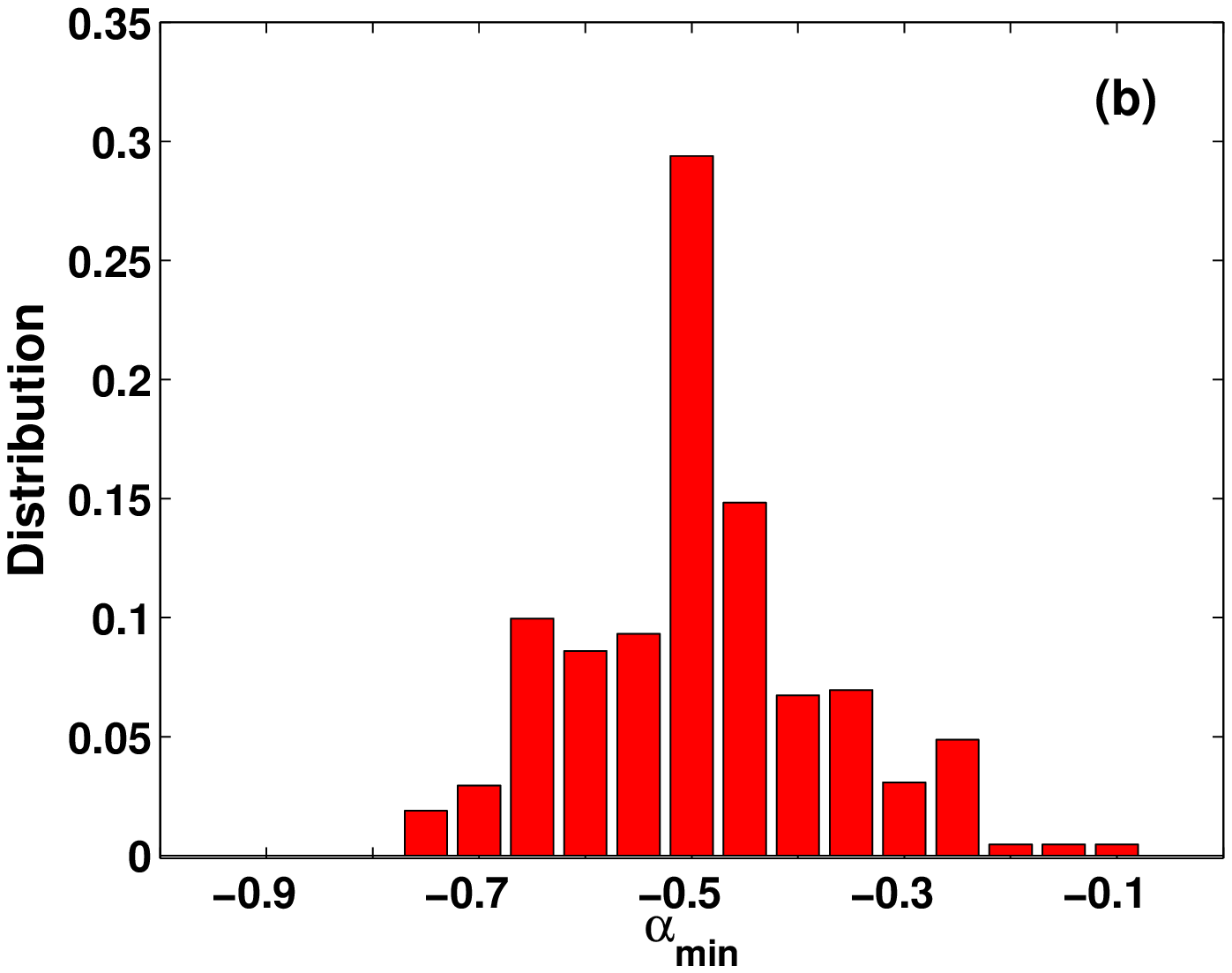,width=0.46\linewidth}
\end{center}
 \caption{(a) Histogram of the minimum slope distribution for all dielectric spectra for the 53 liquids, using subintervals of length of $0.1$ . The number of loss spectra varies widely from liquid to liquid (from 2 to 26), so in order to give all liquids equal weight, each minimum slope value was given the weight $1/N$ if the liquid in question has $N$ spectra included in the analysis. The most frequently observed values of $\am$ are between $-0.45$ and $-0.55$. This implies prevalence of approximate $\sqrt t$ relaxation. (b) Histogram of the same data with subintervals of length $0.05$. Almost a third of the minimum slopes are between $-0.525$ and $-0.475$  \label{fig:alpha}}
\end{figure}

Figure \ref{fig:alpha} shows the minimum slope distribution for the 53 liquids in two histograms of different resolutions. This is the main figure of the paper. The above-discussed limitations, as well as the differing temperature ranges and frequency intervals for the data sets, imply that the number of $\am$ values per liquid varies widely (from 2 to 26). To compensate for this and give equal weight to each liquid, each minimum-slope observation was given the weight $1/N$ where $N$ is the number of spectra for that particular liquid (surviving the data selection criteria).

A priori one would perhaps expect a more or less flat distribution of minimum slopes; nothing in the conventional wisdom indicates that one particular minimum slope should be more likely than another. Our data set, however, show significant prevalence of minimum slopes close to $-1/2$. This corresponds to a prevalence of approximate $\sqrt t$ relaxation of the dielectric relaxation function.

\section{Possible minimum slope correlations}

Assuming that the liquids in the collection are representative of organic glass formers in general, there is something significant with minimum slopes close to $-1/2$. The  obvious question that comes to mind is: How do minimum slopes correlate with other physical quantities? Below we consider six potential correlations.

\subsection{Do minimum slopes correlate with how accurate an inverse power-law fit applies at the inflection point?}

If $\sqrt t$ relaxation were somehow generic for the $\alpha$ process, one would expect that whenever the inflection point tangent gives a particularly good fit, the minimum slope is close to $-1/2$. To look into this we numerically calculated the third-order derivative relative to the first-order derivative of the losses at the inflection point in the usual log-log plot. Defining $H(\log (f))=\log(\eps''(\log(f)))$, the first-order derivative of $H$ with respect to $\log (f)$ at the inflection point frequencyis given by $H^{(1)} = \am$. The second-order derivative $H^{(2)}$ is zero here. Therefore, according to Taylor's theorem a measure of how well the inflection-point tangent approximates the loss, i.e., how well the power-law approximation $\eps'' \propto$ $f^{\alpha_{min}}$ applies, is provided by the ratio between third and first order derivates, $\left|H^{(3)}/ \am \right|$. The smaller this number is, the better is an inverse power-law fit.

\begin{figure}[bt!]
 \begin{center}
 \epsfig{file=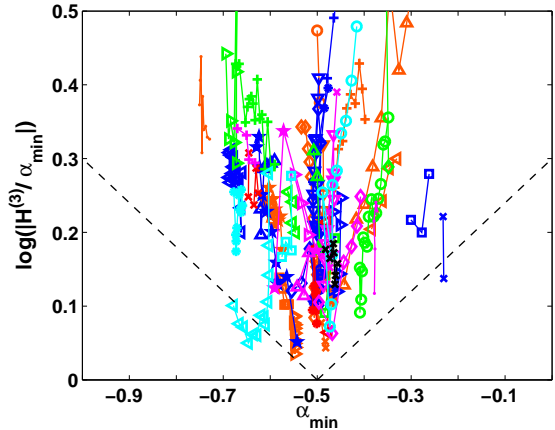,width=0.49\linewidth}
\end{center}
 \caption{A measure of how well the inflection point inverse power-law approximation applies plotted versus minimum slope. The dashed lines are guides for the eye. Every liquid data set is presented with the color and symbol listed in table \ref{tabel}. There is a tendency that liquids where the inverse power-law approximation applies particularly well have minimum slopes close to $-1/2$; thus only two liquids have points below the dashed line. \label{fig:curv}}
\end{figure}

To avoid noise problems we calculated $H^{(3)}$ as the curvature at the minimum of the (previously obtained) graph of the slope as function of frequency. The curvature was calculated by fitting to a second-order polynomial. The number of points in the fitting interval depended on the measured point density and on the symmetry of the neighborhood of this frequency; we used between five and seven points in the fitting intervals. 

Figure \ref{fig:curv} shows $\log\left(\left|H^{(3)}/\am\right|\right)$ versus $\am$ for all spectra. There is no tendency that the power-law approximation works particularly well for liquids with minimum slopes close to $-1/2$. There is, however, the converse tendency indicated by dashed lines that if one requires the power-law approximation to work very well, minimum slopes tend  to be close to $-1/2$. To summarize, Fig. \ref{fig:curv}  confirms a special status associated with liquids with $\am\cong -1/2$.

\subsection{Do minimum slopes and loss-peak frequencies correlate?}

Next we investigate how minimum slopes depend on temperature. If $\am=-1/2$ were generic for the ``pure'' $\alpha$ process, one would expect minimum slopes to converge to this value at low temperatures (still in the metastable equilibrium phase). A convenient way to study $\am$'s temperature dependence is to represent temperature by the loss peak frequency; in this way all liquids are regarded from the same perspective. 

\begin{figure}[bt!]
 \begin{center}
\epsfig{file=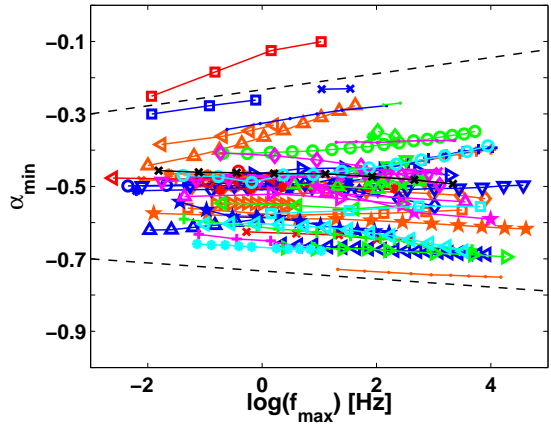,width=0.49\linewidth}
\end{center}
\caption{Minimum slope versus loss peak frequency, the latter being a convenient measure of temperature. There is a  tendency that minimum slopes approach $-1/2$ as temperature is lowered. The dashed lines are drawn as guides to the eye.\label{fig:temp}}
\end{figure}

Figure \ref{fig:temp} shows the results. Minimum slopes are only weakly temperature dependent, but there is a tendency with a few exceptions that liquids with minimum slopes numerically larger than $1/2$ have $|\am|$ decreasing numerically as temperature is lowered, whereas for liquids with minimum slopes numerically smaller than $1/2$, $|\am|$ tends to increase. The dashed lines are drawn to indicate this overall tendency.

Some further notes relating to this figure: Liquids like 2-methyltetrahydrofuran (MTHF, blue {\color{MycBlue} $\nabla$}), DBP (blue {\color{MycBlue} $*$}), DEP (blue {\color{MycBlue} $\circ$}), DOP (orange {\color{MycYell} $\lozenge$}),
 5-polyphenyl-ether (PPE, red {\color{MycRed} $*$}), tetraphenyl-tetramethyl-trisiloxane (DC704 red {\color{MycRed} $\triangleleft$}) and 4-methyl-heptane (4MH, green {\color{MycGreen}  $\star$}) with nearly constant minimum slope close 
to of $-1/2$ all have $\beta$ relaxation loss peaks above $10^5$ Hz. For some glass formers like MMT (blue {\color{MycBlue} $\lozenge$}) $|\am|$ increases above $1/2$, but eventually approaches $1/2$ as temperature is further 
decreased. This presumably reflects the merging of $\alpha$ and low-intensity $\beta$ processes that one observes for scans at temperatures below $T_g$ in Fig. \ref{fig:raw_data1}(l). The same change in $\am$ values is observed for materials with $|\am|>1/2$ like phenolphthalein dimethylether (PDE, cyan {\color{MycCyan} $\triangleleft$}), PG (blue 
{\color{MycBlue} $\triangleleft$}), propylene carbonate (PC, red {\color{MycRed} $\times$}), and nMC (blue {\color{MycBlue} $\vartriangle$}). The dielectric scan of the last liquid nMC in Fig.\ref {fig:raw_data1} (m) shows two secondary processes with times corresponding to frequencies around $100$ Hz and in the interval $0.1-0.01$ Hz, respectively. The loss peak frequencies for the six chosen curves are just above the secondary process ($0.01$ Hz) and $\am$ is decreasing.

In summary, there is a tendency that minimum slopes slowly approach $-1/2$ as temperature is lowered. It would obviously be interesting to have lower temperature observations, but it is not realistic to extend observations to significantly lower temperatures and frequencies while still probing the metastable liquid phase.

\subsection{Do minimum slopes correlate with how well time-temperature superposition applies?}

\begin{figure}[tb!]
\begin{center}
 \epsfig{file=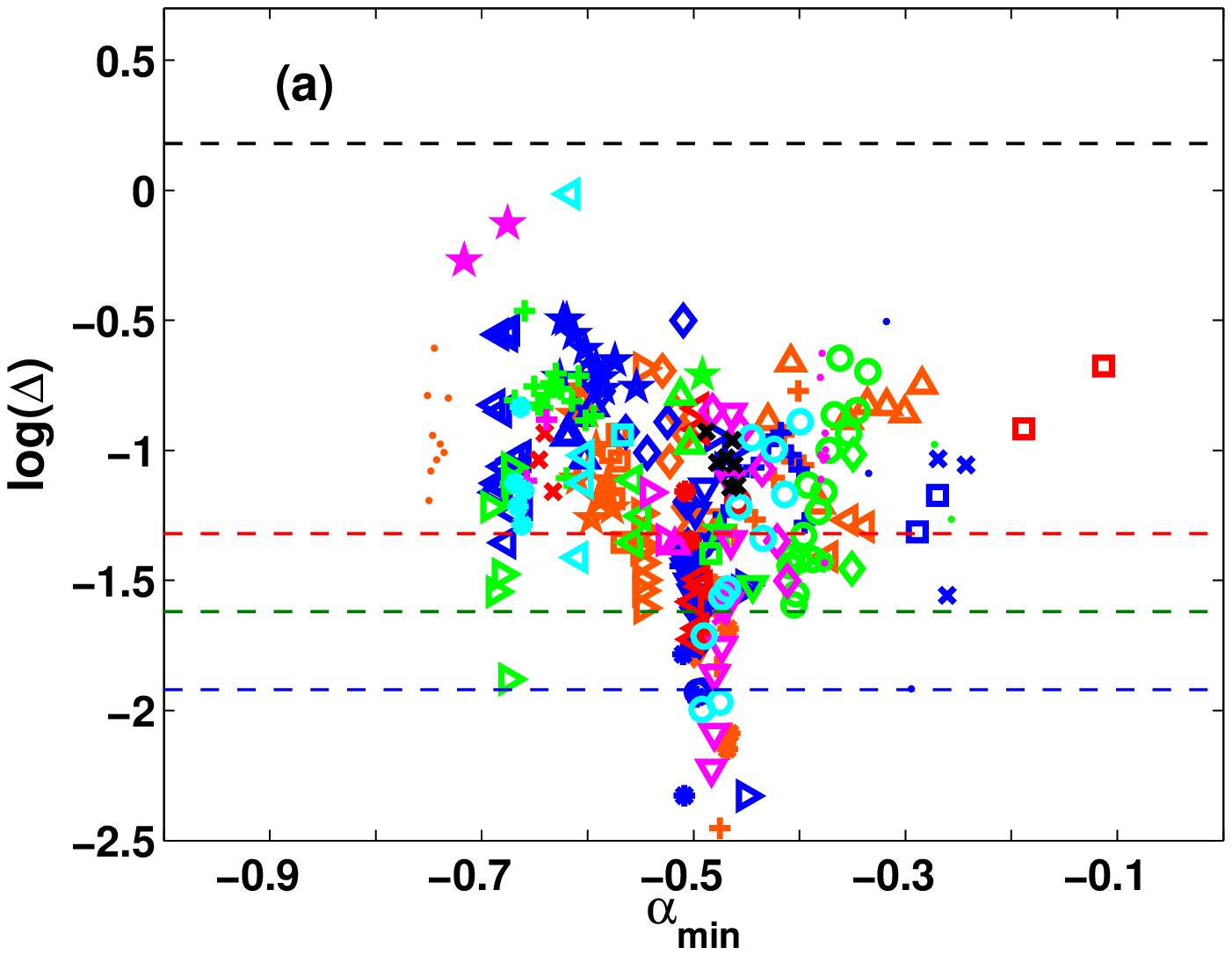 ,width=0.49\linewidth}
 \epsfig{file=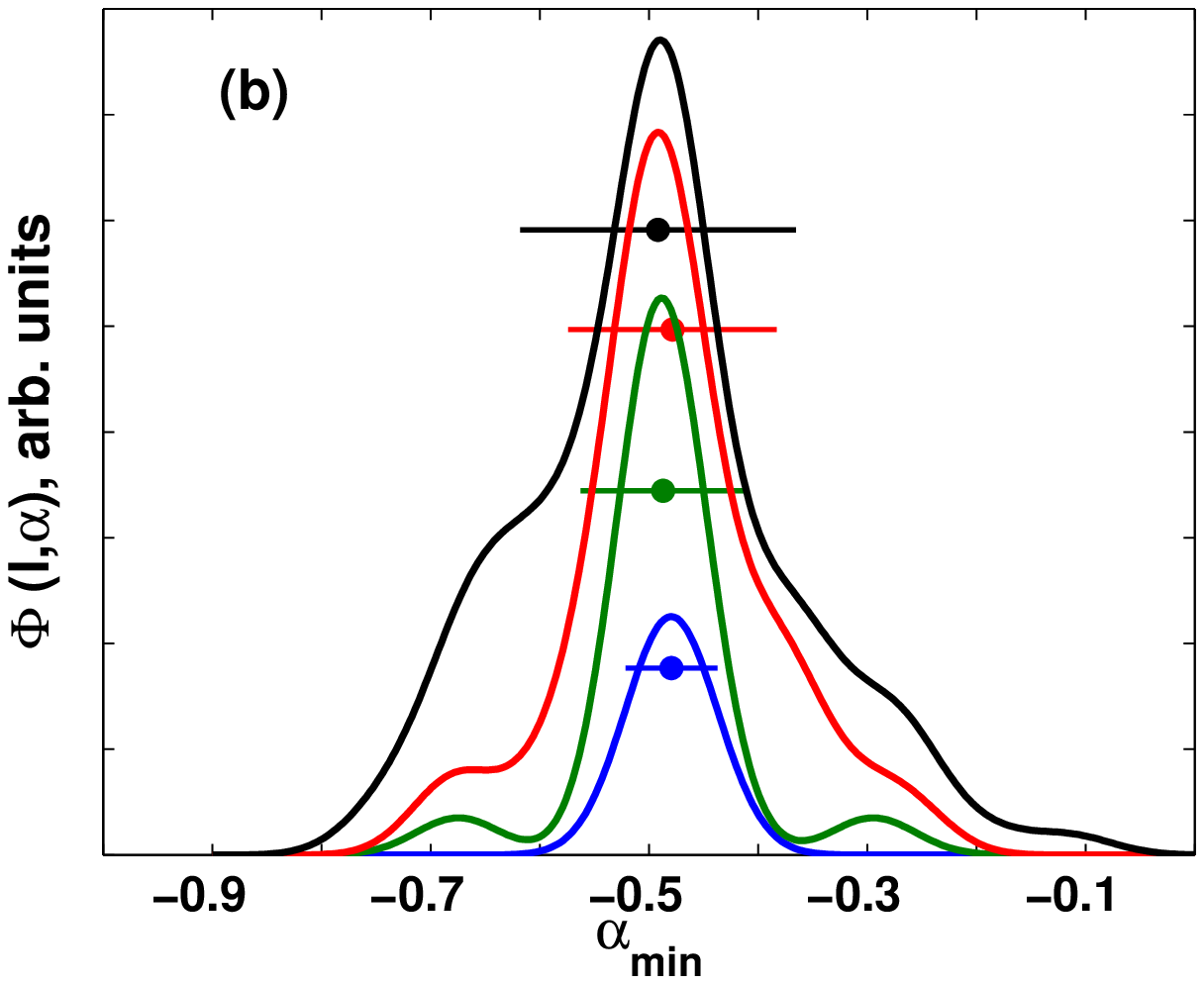,width=0.49\linewidth}
\end{center}
\caption{(a) Time-temperature superposition (TTS) analysis. (a) shows the measure of how well TTS applies, $\log(\Delta)$, plotted versus $\am$. With a few outliers it is seen that the smaller $\log(\Delta)$ is (i.e., the better TTS applies) the more $\am$ tends to $-1/2$. (b) The smoothed distribution 
$\Phi(\alpha,l)$ of the number of measuring points (normalized to the total number of points representing a given liquid) for all liquids with $\log(\Delta)<l$. The levels $l=-1.92 ;  -1.62  ; -1.32 ;   +0.18$ correspond to the colors blue, green, red and black, and are marked with dashed lines in (a). The four dots and vertical lines mark the mean values and variances of $\am$ for the four distributions.\label{fig:TTS}}
\end{figure}

Figure \ref{fig:TTS}(a) shows the TTS measure $\log (\Delta)$ (Eq. (\ref{delta_def})) plotted versus minimum slopes -- in this case the latter were averaged over the two neighboring temperatures involved in defining $\Delta$. The liquids again have varying number of points, so the population of all points on the graph does not give a clear picture of a possible correlation. To compensate for this as was previously done for the minimum slope histogram, in Fig. \ref{fig:TTS}(b) we present the distribution function $\phi$ that gives all liquids equal weight. The distribution function, which is smoothed in this figure, gives information about how many liquids have TTS deviations below a certain level, $l$, for a given value of the $\am$. If $\theta(x)$ is the theta function (unity for positive $x$, zero for negative), $\Lambda=0.003$ is a smoothing parameter, $\alpha_{ij}$ is the minimum slope of $i$-th liquid at the $j$-th temperature in its data series and $\Delta_{ij}$ the corresponding TTS deviation measure, $n=53$ is the total number of liquids, and $N_i$ is the number of spectra of the $i$'th liquid (thus there are $N_i-1$ TTS deviation measures for the liquid), the distribution function is defined as follows: 

\begin{equation}
 \Phi (\am, l)=\frac{1}{n} \sum_{i=1}^{n} \frac{1}{N_{i}-1} \sum_{j=1}^{N_{i}-1} \exp\left(-\frac{\left(\am-\alpha_{ij}\right)^2}{\Lambda}\right)\theta(l-\log(\Delta_{ij}))\,.
\end{equation}

Figure \ref{fig:TTS}(b) gives the function $\phi (\am, l)$ for increasing values of $l$ plotted with blue, green, red, and black, respectively. The corresponding levels $l$ are marked with dashed lines in Fig. \ref{fig:TTS}(a). To the lowest level curve (blue) only the following liquids contribute: $\alpha$-phenyl-o-cresol (PoC, orange  {\color{MycYell} $*$}), polypropylene-glycol 400 (PPG, orange  {\color{MycYell} $+$}), dibutyl phthalate (DBP, blue  {\color{MycBlue} $*$}), APAED (Fig. \ref{fig:raw_data1}(a), magenta {\color{MycMagenda} $\nabla$}), 2MP24D (Fig. \ref{fig:raw_data1}(a) cyan {\color{MycCyan} $\circ$}) and DPGDME (Fig.\ref{fig:raw_data1} (n) , blue {\color{MycBlue} $\triangleright$}). Thus these liquids obey TTS to a very good approximation; they are all  characterized by almost temperature independent $\am\cong -1/2$. %The overall tendency of Fig. \ref{fig:alpha} is reinforced by Fig. \ref{fig:TTS}. Thus minimum slopes around of $-1/2$ dominate, and the distribution narrows around $\am=-1/2$ when the TTS level $l$ decreases. 

In summary, the above confirms the conjecture of Ref. \cite{Olsen01} that liquids accurately obeying TTS have minimum slopes close to $-1/2$. A new observation of the present paper is the general prevalence of $\sqrt t$ relaxation, whether or not TTS applies to a good approximation. %This makes the TTS correlation somewhat less significant for the overall picture.

\subsection{Do minimum slopes correlate with loss peak widths?}

\begin{figure}
 \begin{center}
 \epsfig{file=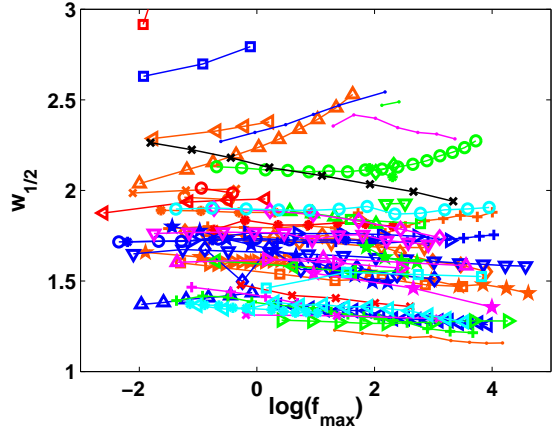,width=0.49\linewidth} 
 \end{center}
 \caption{\label{fig:w_fr} Normalized width $\wh$ plotted versus loss peak frequency, the latter quantity providing a convenient measure of temperature. The width generally changes with temperature and only in some cases becomes almost constant as the  temperature is lowered.}
\end{figure}

The normalized half widths $\wh$ are presented in Fig. \ref{fig:w_fr} versus loss peak frequency. The widths vary between $1.2$ and $3.0$ with the exception of DHIQ (red {\color{MycRed} $\square$}) that has one spectrum with $\wh=4.0$. 

Liquids with almost Debye dissipation have almost same normalized widths ( $1<\wh<1.5$ in Fig. \ref{fig:w_fr}); these liquids are: propylene carbonate (PC, red {\color{MycRed} $\times$}), ethylene glycol (EG, magenta {\color{MycMagenda} $+$}), 1,3PD (orange {\color{MycYell} $\bullet$}), butyronitrile (But, green {\color{MycGreen} $+$}) and dibutylammonium formide (DBAF, green {\color{MycGreen} $\triangleright$}) -- all liquids with strong hydrogen(nitrogen) bonding. To the same group of small-width liquids also belong salol (magenta {\color{MycMagenda} $\times$}) that have minimum slopes close to $-1/2$ and nMC (blue  {\color{MycBlue} $\vartriangle$}) with data points that show that the width narrows as $T\rightarrow T_g$.

\begin{figure}
 \begin{center}
\epsfig{file=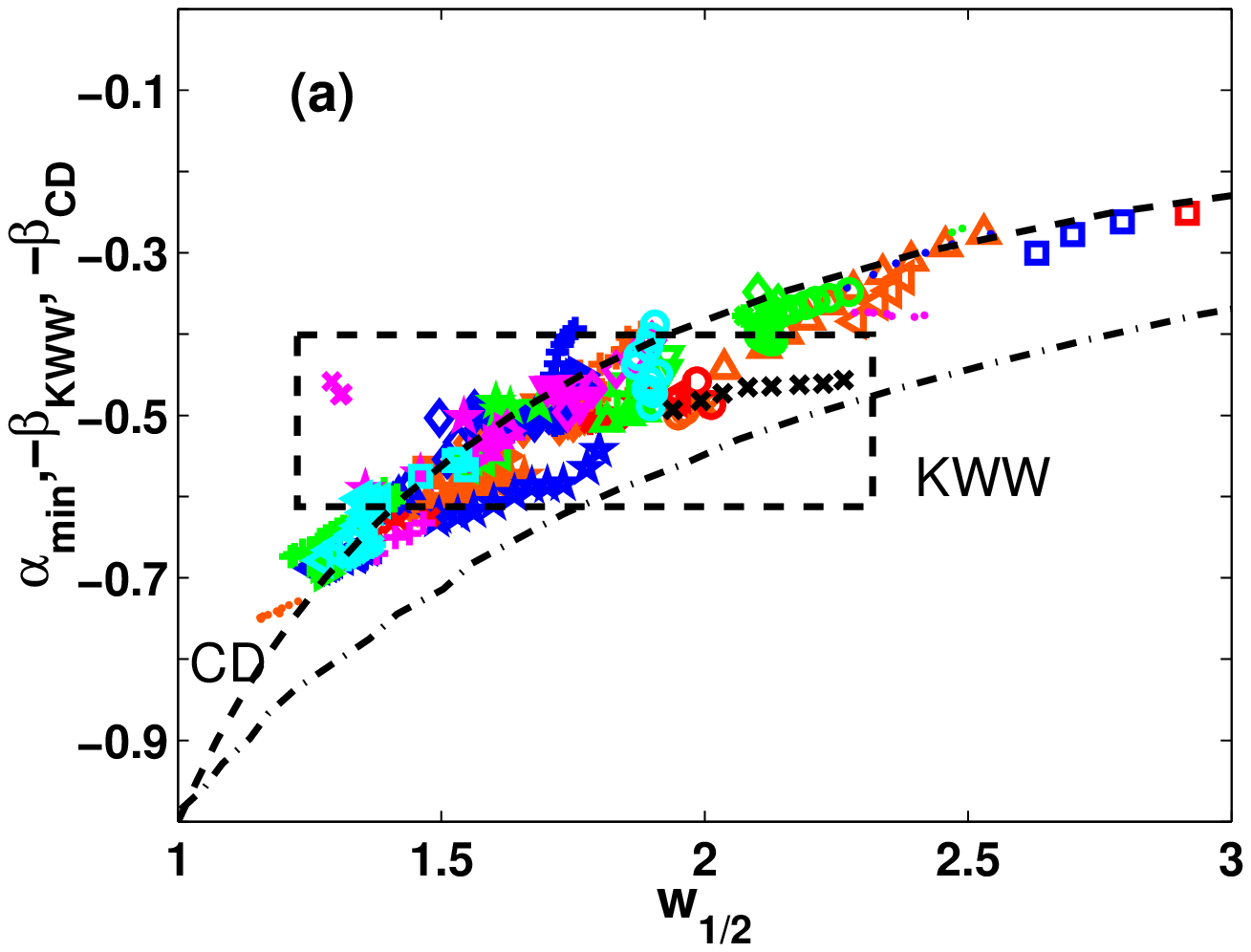,width=0.49\linewidth}
 \epsfig{file=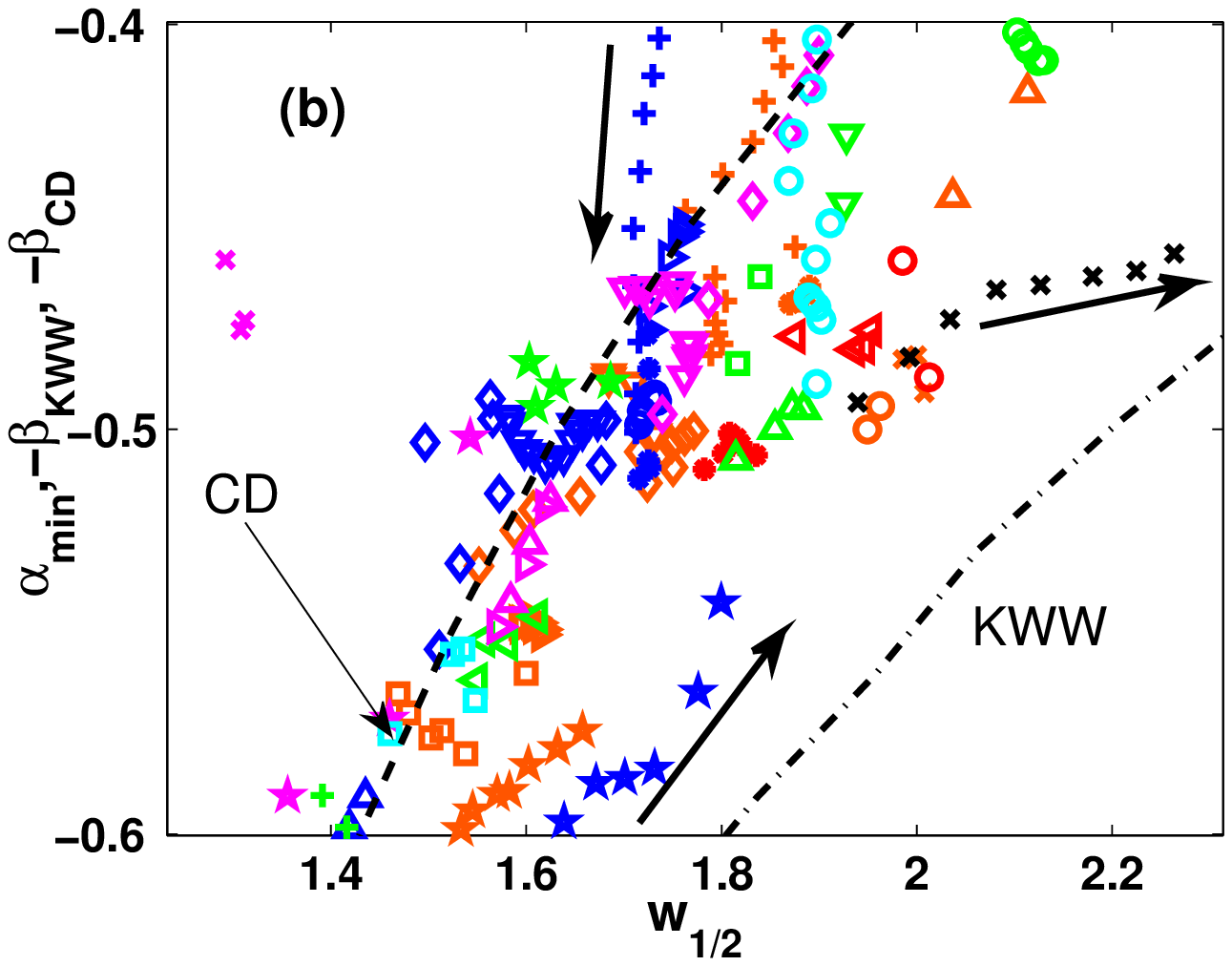,width=0.49\linewidth}
 \end{center}
 \caption{(a) Normalized half  width at half maximum, $\wh$ versus minimum slope $\am$. There is an overall correlation between the two measures, reflecting the fact that a numerically low value of the minimum slope forces the width to be large and vice versa. %An overall tendency is that widths and minimum slopes change with decreasing temperature as marked by the black arrows. 
The dashed-line rectangle frames the zoom-in shown on the plot (b), $-0.6<\am<-0.4$. Here we more clearly see that often minimum slopes vary whereas $\wh$ is nearly constant. In both figures the two black dashed and dash-doted curves give $-\beta_{CD}$, respectively $-\beta_{KWW}$, vs. the corresponding $\wh$. The black arrows indicate the direction of changes as temperature decreases. The values for $\beta_{KWW}$ and $\wh$ for the KWW process are from \cite{hilfer}. \label{fig:w_amin}}
 \end{figure}

Figure  \ref{fig:w_amin} shows $\am$ versus $\wh$. There must be some correlation between $\am$ and $\wh$: If the minimum slope is numerically small, the width must be large and vice versa. In Fig. \ref{fig:w_amin}(a) one indeed finds such a correlation between $\am$ and $\wh$. This is especially apparent for liquids with $\am$ at the boundaries of the $\am$ interval. Thus significant variations of $\wh$ with minimum slope appears for materials with very broad relaxation like sorbitol (blue {\color{MycBlue} $\square$}) and DHIQ (red {\color{MycRed} $\square$}) -- liquids with high-intensity secondary process, as well as Xylitol ({\color{MycBlue} $\bullet$}), 3-methylheptane (3-MH, green {\color{MycGreen} $\bullet$}), TODDA (Fig. \ref{fig:raw_data1}(g), {\color{MycYell} $\triangleleft$}). Sucrose benzoate's (SB, green {\color{MycGreen} $\circ$}) width narrows in the same way, but below some temperature it again begins to grow while the minimum slope gets smaller. This may indicate interference from underlying low-intensity $\beta$ relaxation process (there is an additional well-resolved $\beta$-process above 1 MHz). 

If we focus on minimum-slopes between $-0.4$ and $-0.6$ (Fig. \ref{fig:w_amin}(b)), however, there is a significant spread in the values of normalized widths and no strong correlation between $\wh$ and $\am$. For the glass former MMT (blue {\color{MycBlue}$\lozenge$}) the two quantities are, from some temperature on, almost constant with $\am \in[0.493 ; 0.503]$ and $\wh\in[1.495;1.684]$. Isoeugenol (black {\color{MycBlack} $\times$}) has the same behavior as nMC , the loss peak broadens, but minimum slope is close to $-1/2$.
 Other examples of this are DOP (orange {\color{MycYell} $\lozenge$}), DEP (blue {\color{MycBlue} $\circ$}), and PPE (red  {\color{MycRed} $*$}). For some cases like for DisoBP (blue {\color{MycBlue} $+$}) and DC705 (orange {\color{MycYell} $\circ$}) $\am$ changes significantly while $\wh$ stays almost constant. The reason for this is that $\wh$ does not capture deviations beyond one decade, thus it does not necessarily change when $\alpha$ and $\beta$ processes separate as temperature decrease. In fact, the quantity $\wh$ rarely includes the contributions from around the inflection point that determine the minimum slope.

\begin{figure}
\begin{center}
\epsfig{file=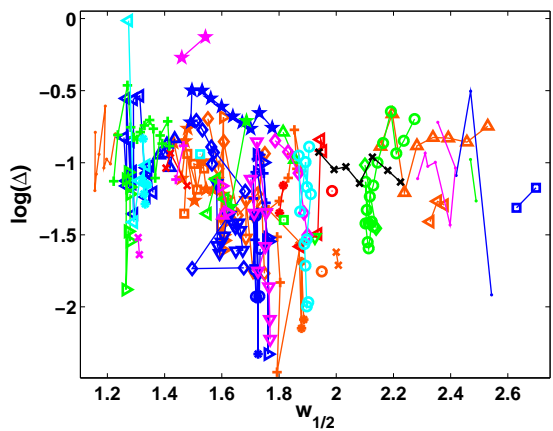,width=0.49\linewidth}
\end{center}
\caption{\label{fig:w_TTS} Normalized width $\wh$ versus the TTS measure $\log(\Delta)$. The figure shows that $\log(\Delta)$ is more noisy than $\wh$, but also more sensitive to shape changes due to temperature decrease, while $\wh$ is in some cases almost constant.}
 \end{figure}

A plot of $\wh$ versus the TTS measure $\log(\Delta)$ is shown in Fig \ref{fig:w_TTS}. We see that $\log(\Delta)$ may be large (and varying) for a given liquid with a fairly constant $\wh$; thus as expected $\log(\Delta)$ is more sensitive than $\wh$ to capturing small changes in the shape of the $\alpha$ process with temperature. Both quantities are affected by noise, of course, and a drawback of $\log(\Delta)$ is that its noise sensitivity has an accumulative character. The ``local'' data noise from the dielectric measuring equipment can be readily seen and noisy data are readily removed from the analysis. Inaccuracies deriving from the sample not being properly thermally equilibrated or from unstable thermal experimental conditions, however, are not so apparent and more difficult to avoid; these are reflected in both measures, but particularly in $\log(\Delta)$.

In summary, there is a mathematically compelling trivial correlation between minimum slopes and loss peak widths, but when one focuses on data sets with $\am\cong -1/2$, a rather broad range of widths is observed, showing that there is little correlation between width and minimum slope for these liquids. Note, incidentally, that this finding emphasizes that single-parameter fits like the stretched exponential or Cole-Davidson are too simple to fit data accurately -- in such fits the width determines the minimum slope and vice versa.

\subsection{Do minimum slopes correlate with how non-Arrhenius the liquid is?}

The two parameters traditionally used to characterize a glass former are its stretching exponent $\beta_{KWW}$ and fragility $m$. The latter measures how much the temperature dependence of the liquid's relaxation time (e.g., inverse loss-peak frequency) deviates from the Arrhenius equation at the glass transition. It generally accepted that the larger the fragility is, the lower is $\beta_{KWW}$ \cite{boehmer1993, wang2007b}; in fact based on experiment a quantitative relation between $m$ and $\beta_{KWW}$ has been suggested \cite{boehmer1994a}. According to this picture all values between $0$ and $1$ for the stretching exponent can occur, depending on the fragility. Since a stretched exponential implies a high-frequency power-law loss varying with frequency as $f^{-\beta_{KWW}}$, from the traditional picture one expects liquids with $\am\cong -1/2$ to have fragilities within a narrow interval.

\begin{figure}
\begin{center}
 \epsfig{file=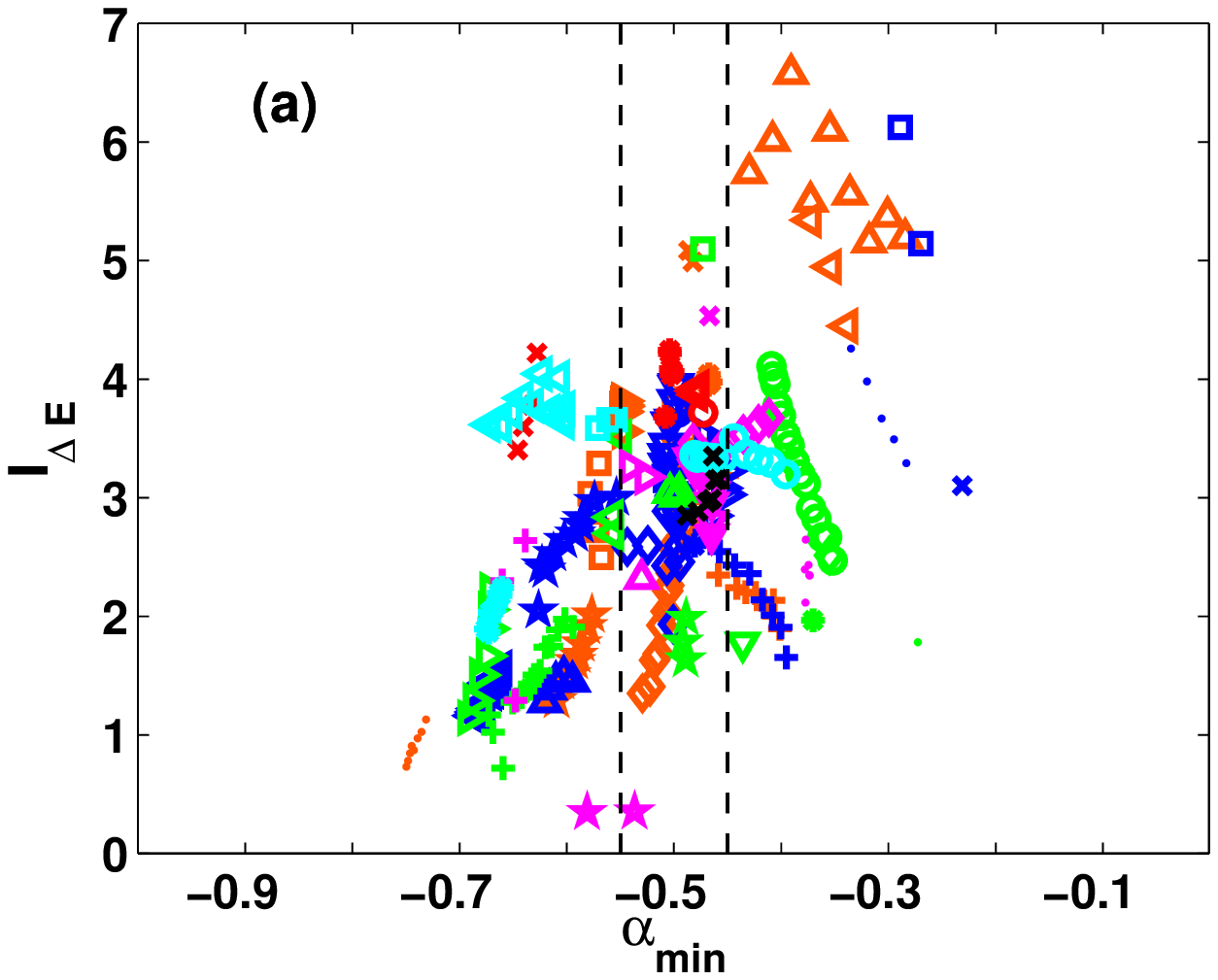,width=0.46\linewidth}
 \epsfig{file=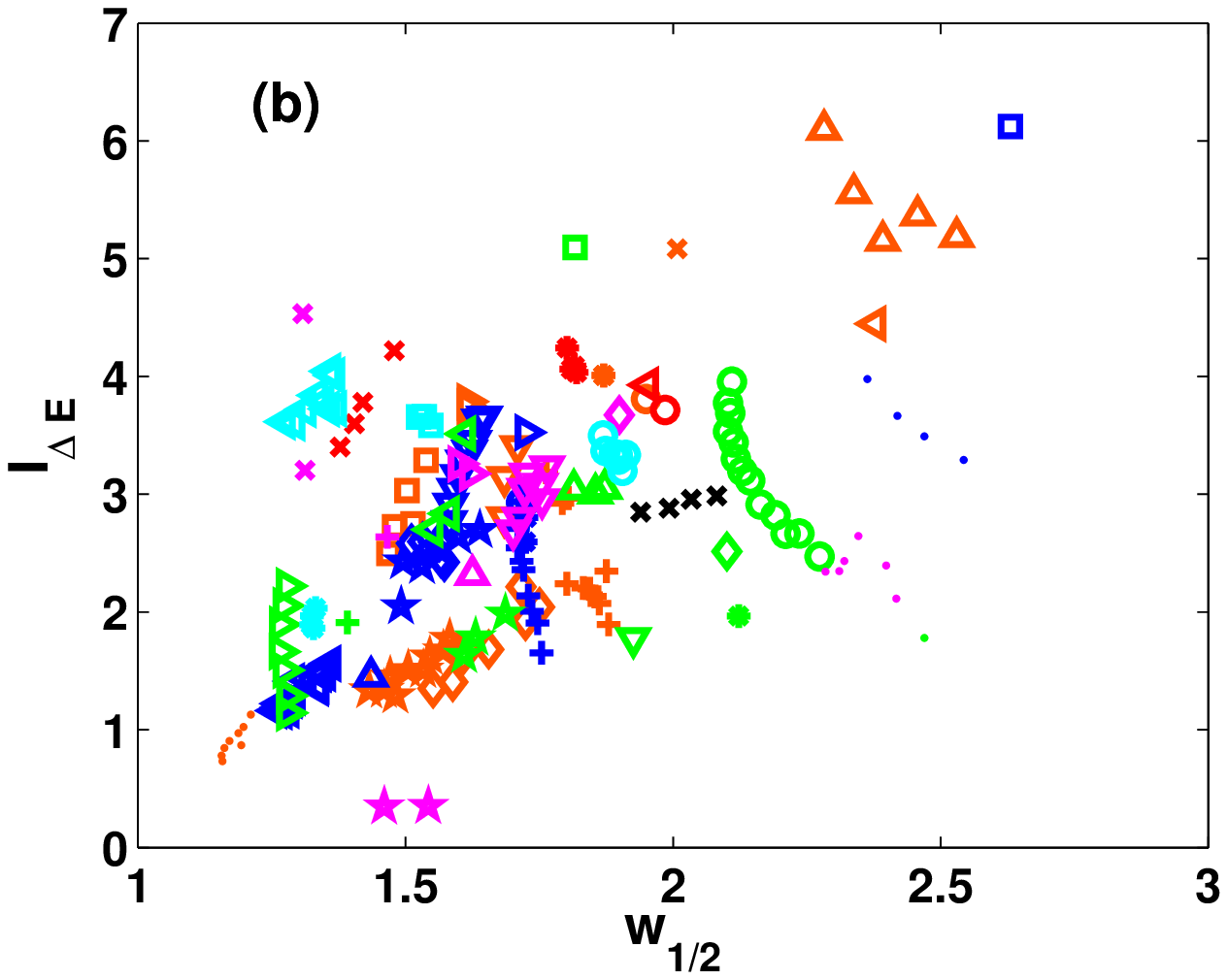,width=0.46\linewidth}
\end{center}
\caption{(a) The activation energy temperature index $I_{\Delta E}$ versus $\am$ for all data sets. The former quantity measures the degree of deviation from Arrhenius temperature dependence of the loss-peak frequency; Arrhenius behavior corresponds to $I_{\Delta E}=0$. The dashed lines  embrace the values between $-0.55$ and $-0.45$. A broad range of non-Arrhenius behaviors is represented among liquids exhibiting approximate $\sqrt t$ relaxation, thus close to $\am=-0.5$ the temperature index varies by a factor of 2.5. In terms of fragility this quantity takes on values from roughly 50 to 125, which is practically the entire span of fragilities of the 53 liquids included in the data analysis. (b) Temperature index $I_{\Delta E}$ versus the normalized width $\wh$ (Eq. (\ref{wdef})), not showing any clear correlation. \label{fig:Boye}}
\end{figure}

We tested the implied correlation between $\am$ and non-Arrhenius behavior by proceeding as follows. As a measure of the degree of non-Arrhenius behavior we used the activation energy temperature index $I_{\Delta E}$ defined \cite{shug1998,  dyre2004, Hecksher08} as follows

\begin{equation}\label{I_def}
I_{\Delta E}(T)\,=\, -\frac{d\ln(\Delta E(T))}{d\ln( T)}\,.
\end{equation}
Here the activation energy $\Delta E(T)$ is defined by writing $\fm(T)=f_0\exp(-\Delta E(T)/k_BT)$ with $f_0=10^{14}$ Hz \cite{Hecksher08}. The temperature index $I_{\Delta E}$ reflects the degree of deviations from Arrhenius behavior at any given temperature. When evaluated at $T_g$ the temperature index relates to $m$ as follows: $m  = 16(I_{\Delta E}(T_g)+1)$ \cite{dyre2004}, where $16=\log(\tau(T_g)/\tau_0)$ if $\tau(T_g)=100$s and $\tau_0=10^{-14}$s. The advantage of using the temperature index for quantifying non-Arrhenius  behavior comes from the fact that the index is defined at any temperature, whereas $m$ is evaluated at  the glass transition temperature and thus formally relates to the liquid's properties only here.

Figure \ref{fig:Boye}(a) plots $I_{\Delta E}$ for all data sets. For liquids exhibiting approximate $\sqrt{t}$ relaxation there is little correlation between the approximate high-frequency power law and the degree of non-Arrhenius behavior. Even the very fragile liquid benzophenone (BP, cyan {\color{MycCyan} $\square$})  ($m=125$ \cite{lunk08}) exhibits approximate $\sqrt t$ relaxation.

For liquids with $\am>-0.4$ we likewise found poor correlation between $\am$ and degree of non-Arrhenius behavior. Thus for DHIQ (red {\color{MycRed} $\square$}) relaxation is characterized by $\am \in [-0.25,-0.10]$), sorbitol (blue {\color{MycBlue} $\square$}), by $\am \in[-0.3, -0.26]$, and salicylsalicylic acid (SSA, blue {\color{MycBlue} $\times$}), by $\am\cong -0.23$, whereas these three liquids have quite different temperature indices (Table 1). For these liquids fragilities reported in the literature are $m= 139$, $m =100$, and $m  = 31(45)$, respectively \cite{remark1}. The lack of clear connection between the shape of the relaxation and the fragility is also clear in the plot $I_{\Delta E}$ versus $\wh$ in Figure \ref{fig:Boye} (b). 

To summarize, liquids with approximate $\sqrt t$ relaxation exhibit a wide range of temperature indices (fragilities); there is no obvious correlation between the degree of non-Arrhenius temperature dependence of the loss peak frequency and the high-frequency decay of the loss.

\subsection{Do minimum slopes correlate with dissipation magnitudes?}

As a measure of dielectric strength one would prefer the overall loss $\Delta\eps$, but since this quantity may be difficult to determine accurately we instead quantify the strength by the maximum loss. These two quantities are only strictly proportional for liquids with same relaxation function, of course, but this fact is not important here because the dielectric strengths span more than four decades.

\begin{figure}
\begin{center}
 \epsfig{file=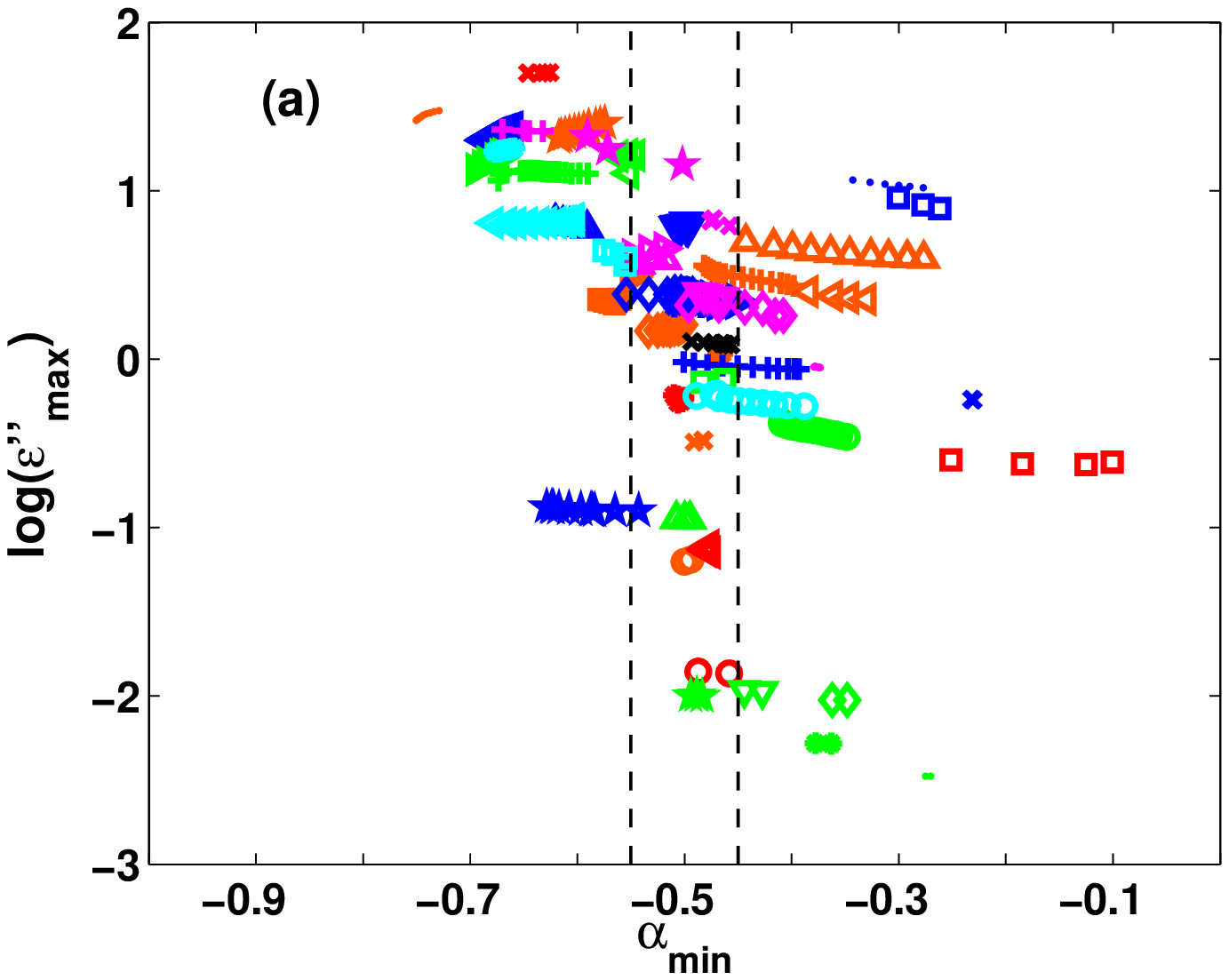,width=0.49\linewidth}
 \epsfig{file=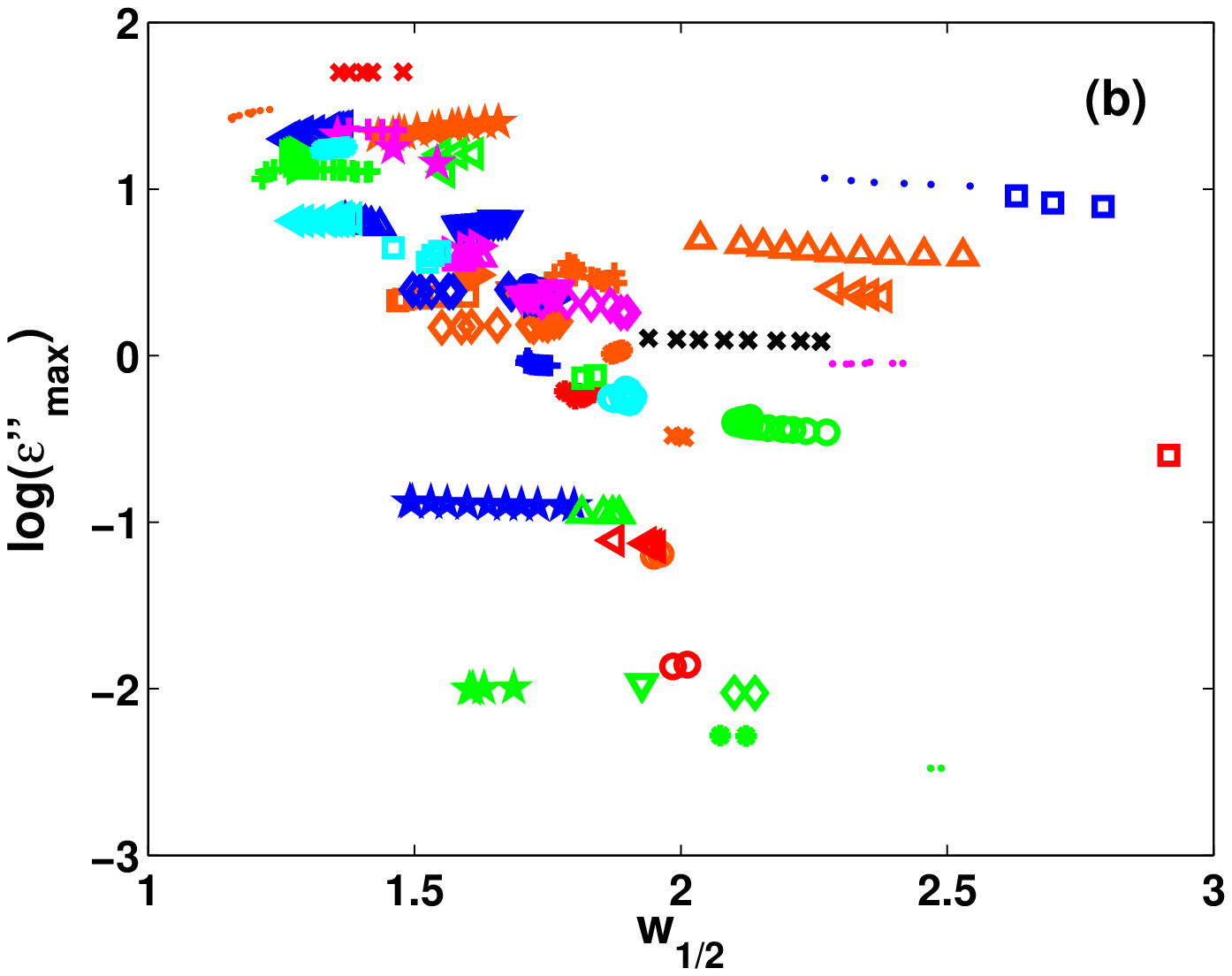,width=0.49\linewidth}
\end{center}
\caption{(a) Maximum dielectric loss $\eps''_{max}$ versus $\am$ for all data sets. The liquids between the two dashed lines marking the interval $-0.55 < \am < -0.45$ have dielectric losses varying by more than a factor of 1,000. Large-loss liquids have minimum slopes that are numerically larger than $1/2$; these liquids consistently disobey approximate $\sqrt{t}$ relaxation. (b) Maximum dielectric loss plotted versus width $\wh$. Glass formers with large dielectric loss consistently tend to be more Debye like as expected from (a). \label{fig:cor}}
\end{figure}

As can be seen from Fig. \ref{fig:cor} (a) there is little overall correlation between having $\sqrt{t}$ relaxation and the value of the maximum loss $\log(\eps''_{max})$. However, liquids with large dielectric strength like PDE (cyan {\color{MycCyan} $\square$}), PG (blue {\color{MycBlue} $\triangleleft$}), PC (red {\color{MycRed} $\times$}), EG (magenta {\color{MycMagenda} $+$}), 1,3PD (orange {\color{MycYell} $\bullet$}),  butyronitrile (green {\color{MycGreen} $+$}), and DBAF (green {\color{MycGreen}$\triangleright$}) consistently show minimum slopes that are numerically larger than $1/2$. The corresponding $\am$ values are only weakly temperature dependent, which agrees with results for other hydrogen-bonding systems \cite{defrancesca}. Liquids with $|\am| > 0.65$ tend to have Kirkwood correlation factors \cite{kirkwood} significantly larger than unity, reflecting strong correlations between the motions of different dipoles. Higher Kirkwood correlation factors mean longer-range orientational and dynamical correlations, leading to spatial averaging of what might otherwise still be $\am=-1/2$ behavior (for Kirkwood correlation factors going to infinity one expects an approach to Debye relaxation because of the increasingly large degree of cooperativity). Figure \ref{fig:cor} (b) shows loss-peak strength versus width. There is a clear tendency that large-strength liquids are more Debye like.

To summarize, liquids with approximate $\sqrt t$ relaxation span a wide range of dielectric losses. There is little overall correlation between loss strength and minimum slope. Liquids with large loss strengths, though, clearly have $|\am|>1/2$.

\section{Conclusions}

The data compiled in this study suggest that -- with the exception of large-loss liquids -- $\sqrt t$ relaxation is generic to the $\alpha$ process of glass-forming liquids. This conclusion is not only based on the observed prevalence of $\sqrt{t}$ relaxation (Fig. 4), but also on our findings that:
\begin{itemize}
\item The better an inverse power law describes the high-frequency loss, the closer are minimum slopes to $-1/2$ (Fig. \ref{fig:curv}).  
\item The lower temperature is, the closer are minimum slopes to $-1/2$ (Fig. \ref{fig:temp}).
\item The better TTS applies, the closer are minimum slopes to $-1/2$ (Fig. \ref{fig:TTS}).
\end{itemize}

Intuitively, one would expect that interference from $\beta$ processes can only explain minimum slopes that are numerically smaller than $1/2$. From measurements on liquids with a well-defined $\beta$ process in the kHz range, however, we and other groups have repeatedly found that when the liquid is heated above the $\alpha\beta$ merging temperature, the high-frequency decay of the merged process has a minimum slope that is numerically larger than $1/2$ (and eventually converges to one upon further heating). Thus, since whenever there are very low-lying beta processes the liquid is above the $\alpha\beta$ merging temperature, $|\am| > 1/2$ might well occur at the lowest attainable temperatures for some liquids. ``Genuine'' $\am=-1/2$ behavior only appears when the system is significantly below the merging temperature, a situation that for several liquids is experimentally out of reach.

Liquids exhibiting approximate $\sqrt t$ relaxation have no particular loss peak widths, temperature indices (fragilities), or loss magnitudes. %Even TTS does not correlate obviously to exhibiting approximate $\sqrt t$ relaxation, because the latter not only dominates for liquids that accurately obey TTS (as previously conjectured \cite{Olsen01}), but in fact for the entire data collection.

A potential weakness of the analysis is that no objective criteria can be given for the selection of liquids included in the analysis. Thus there is the danger of unknowingly having a bias in the data. The data were gathered from leading groups in the field and supplemented by new or previously unpublished measurements. As detailed above, several data sets were discarded in the process because of having too much noise or other problems. The fact that we cannot report objective liquid selection criteria for the initial data pool makes, the analysis should be suplemented with data for other liquids before a firm conclusion can be drawn that approximate $\sqrt t$ relaxation is generic to the $\alpha$ relaxation process. 

If $\sqrt{t}$ relaxation is confirmed as being generic for the $\alpha$ process (excluding high-loss liquids), the dynamics of glass-forming organic liquids is simpler than presently generally believed. That presents an important challenge to theory -- although it should be noted again that there are already several theories predicting this \cite{glarum,doremus,bordewijk,isak,Barlow,montrose1970,Dyre05a}.

\acknowledgments

 For kindly providing data to this study we are indebted to: S. Benkhof, T. Blochowicz,  L. F. del Castillo, R. Diaz-Calleja, L.-T. Duong, K. Duvvuri, G. Eska, C. Gainaru, A. Garcia-Bernabe, S. Hensel- Bielowka, W. Huang, N. Ito,  E. Kaminska, M. K\"{o}hler, A. Kudlik, A. Loidl, P. Lunkenheimer, D. V. Matyushov, M. Mierzwa, P. Medick, K. L. Ngai,  V. N. Novikov, M. Paluch, S. Pawlus, L. C. Pardo, S. Putselyk, E. L. Quitevis, J. R. Rajian, J. R. Rajesh,  A. Rivera, E. A. R\"{o}ssler, M. J. Sanchis, S. Shahriari, N.V. Surovtsev, C. Tschirwitz, L.-M. Wang, J. Wiedersich, and J. Ziolo. Furthermore we thank R. Hilfer for providing half-loss widths for the stretched exponential function.

The centre for viscous liquid dynamics ``Glass and Time'' is sponsored by the Danish National Research Foundation (DNRF).

\clearpage
\vspace{2cm}
\begin{footnotesize}
 
\begin{singlespacing} 

\begin{longtable}{l|l|c|c|c|c|c|c}
\caption{List of all liquids studied providing relevant references and information such as glass transition temperature, $T_g$, and intervals for quantities characterizing the data: the activation energy temperature index,  $ I_{\Delta E}\equiv |d\ln(\Delta E)/d\ln (T)|$; temperature $T$; maximum dielectric loss i$\log(\varepsilon''_{max})$; and minimum slopes of the log-log plot of the loss $| \alpha_{ min}|$.The data listed bellow can be obtained from the ``Glass and Time: Data repository'', found online at http://glass.ruc.dk/data. \label{tabel}}\\

\hline
 \multicolumn{1}{l|}{\textbf{Liquid}} &
\multicolumn{1}{l|}{\textbf{Abbrevi-}} &
\multicolumn{1}{c|}{\textbf{${\mathbf T_g}$ (K)}}&
\multicolumn{4}{c|}{\textbf{Intervals}} &
\multicolumn{1}{c} \textbf{Symbol}\\
\cline{4-7}& \textbf{ation}		  						&				 & ${ { I_{\Delta E}}}$  & $ { T}$ (K) & ${{log \varepsilon''_{max} }}$ & ${| \alpha_{min}|} $& \textbf{and ref.}
 \\ \hline 
\endfirsthead

\multicolumn{8}{c} {{\bfseries \tablename\ \thetable{} -- continued from previous page}} \\
\hline \multicolumn{1}{l|}{\textbf{Liquid}} &
\multicolumn{1}{l|}{\textbf{Abbrevi-}} &
\multicolumn{1}{c|}{\textbf{${\mathbf T_g}$ (K)}}&
\multicolumn{4}{c|}{\textbf{Intervals}} &
\multicolumn{1}{c} \textbf{Symbol}\\
\cline{4-7}& \textbf{ation}		  						&				 & ${{ I_{\Delta E}}}$  & $ {T}$ (K) & ${ {log \varepsilon ''_{max} }}$ & ${  |\alpha_{min}|} $& \textbf{and ref.}
 \\ \hline 
\endhead
 \hline
\endfoot
\hline
 
1,1'-bis&BPC& 212&$3.67;3.67$ & $338;362$& $0.258;0.321$ & $0.41 ; 0.5$&  {\color{MycMagenda} $\lozenge$}\\
(methoxyphenyl)- &&&&&&&\\
cyclohexane &&&&&&&\cite{ngai}\\  \hline

1,2-propanediol &PG&$168$    & $1.16;1.56]$ & $180;205$& $1.3;1.375$ & $0.66 ; 0.69$&  {\color{MycBlue} $\triangleleft$} \\ &&\cite{tpgdpg}&&&&&this work\\\hline

1,3-propane&13PD&$167$  & $0.73;1.13$ & $165;189$& $1.419;1.477$ & $0.73 ; 0.75$&  {\color{MycYell} $\bullet$} \\  diol && \cite{RA1998}  &&&&& this work \\ \hline

2,3-dimethyl-&2,3-DMP&$87.5$& $1.78;1.78$ & $98;99$& $-1.971;-1.967$ & $0.43 ; 0.44$&  {\color{MycGreen} $\nabla$} \\ pentane &&&&&&& \cite{Shahriari2004}\\ \hline

2,3-epoxy-&23EPPPE& 193 & $3.74;3.79$ & $196;200$& $0.483;0.522$ & $0.55 ; 0.55$&  {\color{MycYell} $\triangleright$}\\ propyl-&&&&&&& \\ 
 phenylether &&&&&&& this work\\\hline

2,4,6-trimethyl-&246TMH&$123$ & $2.51;2.51$ & $134;135$& $-2.025;-2.024$ & $0.35 ; 0.36$&  {\color{MycGreen} $\lozenge$}\\heptane &&&&&&&  \cite{Shahriari2004}\\ \hline

2-methyl-&2MP24D& 187 & $3.2;3.5$ & $210;232$& $-0.28;-0.202$ & $0.39 ; 0.49$&  {\color{MycCyan} $\circ$} \\pentane-  &&&&&&&\\2,4-diol  &&&&&&&this work\\ \hline

2-methyl-&MTHF&	$91$& $2.77;3.66$ & $91;103$& $0.776;0.815$ & $0.5 ; 0.51$&  {\color{MycBlue} $\nabla$} \\tetrahydrofuran  &&&&&&&this work\\ \hline

2-phenyl-&APED&222& $2.69;3.23$ & $220;240$& $0.357;0.397$ & $0.46 ; 0.49$&  {\color{MycMagenda} $\nabla$} \\  5-acetomethyl-  &&\cite{number} &&&&&\\5-ethyl- 1,3-  &&&&&&&\\dioxocyclo- &&&&&&&this work\\ hexane&&&&&&&\\ \hline

2-picoline&2pic&	$130$& $3.17;3.26$ & $135;141$& $0.618;0.658$ & $0.52 ; 0.55$&  {\color{MycMagenda} $\triangleright$} \cite{Gai2005}\\ \hline

3-fluoro&3FA&$172$& $5.1;5.1$ & $235;239$& $-0.135;-0.121$ & $0.46 ; 0.48$&  {\color{MycGreen} $\square$}\\
-aniline &&\cite{wang2007b}&&&&& \cite{wied1999}\\\hline

3-methyl-&3MH&$97$&$1.78;1.78$ & $109;110$& $-2.477;-2.477$ & $0.27 ; 0.27$&  {\color{MycGreen} $\bullet$}\\ heptane&&&&&&& \cite{Shahriari2004}\\ \hline

3-methyl-&3MP&$79$ & $1.97;1.97$ & $88;89$& $-2.283;-2.281$ & $0.36 ; 0.38$&  {\color{MycGreen} $*$}\\ pentane&&&&&&& \cite{Shahriari2004}\\\hline

4-methyl-&4MH& 99 & $1.63;1.98$& $111;114$& $-2.004;-1.995$ & $0.48 ; 0.49$&  {\color{MycGreen}  $\star$} \\heptane  &&&&&&&\cite{Shahriari2004}\\ \hline

4-tertbuthyl-&4TBP&$166$ & $2.32;13.79$ & $164;177$& $0.566;0.602$ & $0.52 ; 0.54$&  {\color{MycMagenda} $\vartriangle$} \\ pyridine &&&&&&&\cite{bloch2006}\\ \hline

4,7,10-&TOTDD&$108$ & $4.45;4.45$ & $177;181$& $0.356;0.401$ & $0.33 ; 0.38$&  {\color{MycYell} $\triangleleft$}\\trioxatridecane-&&&&&&& \\1,13- diamine &&&&&&& this work\\ \hline

5-polyphenyl-&PPE& $248$ & $4.04;4.24$ & $252;264$& $-0.258;-0.214$ & $0.5 ; 0.51$&  {\color{MycRed} $*$} \\ether&&&&&&& \cite{Jacobsen05}\\ \hline

$\alpha$-phenyl-o-&PoC& $219$ & $4.01;4.01$ & $220;228$& $0.011;0.032$ & $0.46 ; 0.47$&  {\color{MycYell} $*$} 
\\cresol&&&&&&& this work\\ \hline

benzophenone&BP& $212$& $3.59;3.66$ & $215;230$& $0.56;0.647$ & $0.55 ; 0.58$&  {\color{MycCyan} $\square$} \cite{lunk08} \\ \hline

biphenyl-2yl-&BP2BF&$210$ & $1.86;2.03$ & $190;200$& $1.232;1.253$ & $0.66 ; 0.68$&  {\color{MycCyan} $*$} \\ isobutylate  && \cite{number}&&&&& this work\\\hline

butyronitrile&But& $95$&  $1.91;1.91$ & $98;116$& $1.061;1.121$ & $0.59 ; 0.67$&  {\color{MycGreen} $+$} \cite{ito2006}\\ \hline

decahydro-&DHIQ&$180$& $7.13;7.13$ & $180;185$& $-0.626;-0.599$ & $0.1 ; 0.25$&  {\color{MycRed} $\square$} \\isoquinoline&&\cite{richert2003}&&&&&\cite{richert2003, Jacobsen05}\\ \hline

dibutyl-&DBAF& $153$    & $1.14;2.22$ & $162;185$& $1.127;1.218$ & $0.67 ; 0.69$&  {\color{MycGreen} $\triangleright$} \\ammonium- &&&&&&& \\ formide &&&&&&&\cite{ito2006b}\\ \hline

dibutyl &DBP& 	$177$ & $2.6;3.07$ & $178;192$& $0.301;0.348$ & $0.48 ; 0.51$&  {\color{MycBlue} $*$} \\ phthalate&&&&&&&this work\\ \hline

di-\textit{iso}-butyl&DisoBP& 191  & $1.65;2.94$ & $201;221$& $-0.06;-0.016$ & $0.39 ; 0.5$&  {\color{MycBlue} $+$} \\  phthalate&&\cite{wang2007b}&&&&&this work\\\hline

dicyclohexyl&DCMMS &$220$& $2.8;3.41$ & $224;240$& $0.381;0.411$ & $0.49 ; 0.5$&  {\color{MycYell} $\nabla$} \\ -methyl-2- &&&&&&&\\methyl- &&&&&&&\cite{Diaz}\\succinate &&&&&&&\\\hline

dicyclohexyl &DCHMS& 222 & $2.11;2.64$ & $218;230$& $-0.05;-0.041$ & $0.37 ; 0.38$&  {\color{MycMagenda} $\bullet$} \\ -2-methyl- && \cite{number}&&&&& this work\\ succinate &&&&&&&\\ \hline

diethyl &DEP&187& $2.93;2.93$ & $183;192$& $0.375;0.412$ & $0.49 ; 0.5$&  {\color{MycBlue} $\circ$} \\phthalate&& \cite{wang2007b}&&&&&this work\\ \hline 

diglycidyl-&ER& 259& $3.67;3.67$ & $338;362$& $0.258;0.321$ & $0.41 ; 0.5$&  {\color{MycMagenda} $\lozenge$} \\ether &&&&&&&\\of bisphenol A &&&&&&&\cite{epoxy}\\ (epoxy-resin)&&&&&&&\\\hline

dioctyl &DOP& $189$& $1.35;2.21$ & $190;220$& $0.168;0.205$ & $0.5 ; 0.53$&  {\color{MycYell} $\lozenge$} \\ phthalate&& \cite{DOP}&&&&&this work\\\hline

dipropylene-&DPGDME&137 & $3.52;3.52$ & $139;151$& $0.327;0.373$ & $0.45 ; 0.48$&  {\color{MycBlue} $\triangleright$} \\dimethyl- && \cite{Paluch} &&&&&this work\\glycol- &&&&&&&\\dimethylether &&&&&&&\\ \hline

ethylene glycol&EG&$152$  & $2.64;2.64$ & $158;165$& $1.354;1.364$ & $0.63 ; 0.67$&  {\color{MycMagenda} $+$}  \\glycol&&&&&&&\cite{EG}\\ \hline

glycerol&Gly&$193$& $1.29;1.77$ & $192;236$& $1.317;1.401$ & $0.57 ; 0.62$&  {\color{MycYell}  $\star$} \cite{Olsen01}\\ &&\cite{RA1998} &&&&&\\ \hline

isoeugenol&&$220$  & $2.85;2.99$ & $225;248$& $0.085;0.104$ & $0.46 ; 0.49$&  {\color{MycBlack} $\times$}\\&&&&&&&this work \\ \hline

isopropyl- &Cumene& $126$& $3.01;3.05$ & $135;139$& $-0.951;-0.948$ & $0.49 ; 0.51$&  {\color{MycGreen} $\vartriangle$}\\ benzene&&&&&&& \\(cumene) &&&&&&&this work \\\hline

methyl-m-&MMT&$165$  	& $2.42;2.6$ & $173;189$& $0.371;0.397$ & $0.49 ; 0.55$&  {\color{MycBlue} $\lozenge$} \\toluate &&&&&&&this work\\ \hline

n-$\varepsilon$-methyl-&nMC  & 172 & $1.45;1.45$ & $186;196$& $0.778;0.816$ & $0.59 ; 0.62$&  {\color{MycBlue} $\vartriangle$} \\caprolactam&&\cite{hansen1997} &&&&&this work\\ \hline

n-propyl-&nPB& $122$ & $2.05;2.7$ & $127;137$& $-0.902;-0.878$ & $0.54 ; 0.63$&  {\color{MycBlue}  $\star$} \\ benzene&&\cite{wang2007b} &&&&&this work\\\hline

phenol-&PDE&295 & $3.61;4.04$ & $301;325$& $0.808;0.833$ & $0.6 ; 0.68$&  {\color{MycCyan} $\triangleleft$} \\ phthalein-&&\cite{shar2007} &&&&&\\dimethylether &&&&&&&\cite{Hensel-Bielowka2002} \\ \hline

phenylsalicate &Salol& 215& $3.2;4.53$ & $177;187$& $0.793;0.834$ & $0.46 ; 0.48$&  {\color{MycMagenda} $\times$} \\(salol)&& \cite{hatasea}&&&&&\cite{Gai2005}\\ \hline

polypropylene-&PPG& $73$& $1.9;3.19$ & $200;226$& $0.436;0.556$ & $0.4 ; 0.48$&  {\color{MycYell} $+$}\\glycol 400&&&&&&& \cite{Olsen01}\\ \hline

propylene &PC & $160$& $3.4;4.22$ & $162;170$& $1.699;1.703$ & $0.63 ; 0.65$&  {\color{MycRed} $\times$}\\ carbonate&&&&&&& \cite{Wang2002PC}\\\hline

salicyl- &SSA& $279$& $3.1;3.1$ & $305;308$& $-0.243;-0.238$ & $0.23 ; 0.23$&  {\color{MycBlue} $\times$} \\ salicylic acid && \cite{SSA}&&&&&this work\\\hline

sorbitol&Sor&268& $6.12;6.12$ & $268;273$& $0.895;0.959$ & $0.26 ; 0.3$&  {\color{MycBlue} $\square$} \\ &&\cite{wang2007b} &&&&&this work \\\hline

sucrose-&SB&337 & $2.47;3.96$ & $343;373$& $-0.461;-0.373$ & $0.35 ; 0.41$&  {\color{MycGreen} $\circ$}\\ benzoate&&&&&&& \cite{Raj06}\\\hline

tetraphenyl-&DC704&211 & $3.93;3.93$ & $211;219$& $-1.148;-1.109$ & $0.48 ; 0.48$&  {\color{MycRed} $\triangleleft$} \\ tetramethyl- &&&&&&&\cite{Jacobsen05}\\ trisiloxane &&&&&&&\\ \hline

tricresyl-&TCP&$211$ & $2.5;3.29$ & $214;236$& $0.33;0.356$ & $0.56 ; 0.58$&  {\color{MycYell} $\square$} \\
phosphate &&&&&&& \cite{Hecksher08}\\ \hline

trimethyl-&DC705& $230$ & $3.81;3.81$ & $233;235$& $-1.203;-1.191$ & $0.49 ; 0.5$&  {\color{MycYell} $\circ$} \\pentaphenyl&&&&&&&\\  trisiloxane&&&&&&&this work \\\hline

trimethyl &3MPh& 136& $2.7;3.51$ & $141;150$& $1.104;1.214$ & $0.55 ; 0.56$&  {\color{MycGreen} $\triangleleft$}\\  phosphate&&&&&&& \cite{bloch2006}\\ \hline

 triphenyl&TPP& 204 & $5.08;5.08$ & $204;208$& $-0.493;-0.479$ & $0.48 ; 0.49$&  {\color{MycYell} $\times$}\\ phosphite&&&&&&& \cite{Olsen01}\\ \hline

triphenyl-&TPE&249& $3.72;3.72$ & $256;258$& $-1.866;-1.856$ & $0.46 ; 0.49$&  {\color{MycRed} $\circ$} \\ ethylene&& \cite{NJmaster} &&&&&\cite{Jacobsen05}\\ \hline

toluene-&TolPyr&$123$&   $5.16;6.1$ & $126;131$& $0.597;0.698$ & $0.28 ; 0.44$&  {\color{MycYell} $\vartriangle$}\\ pyridine&& \cite{Olsen00}&&&&&\\ mixture&&&&&&& \cite{Olsen98}\\  \hline

xylitol&Xylitol& $248$ &  $3.29;3.98$ & $254;266$& $1.019;1.065$ & $0.28 ; 0.34$&  {\color{MycBlue} $\bullet$} \\ &&\cite{wang2007b} &&&&&this work \\\hline 

\end{longtable}

\end{singlespacing}

\end{footnotesize}


\begin{thebibliography}{Bibliography}
\bibitem{note1} There may be several secondary $\beta$ processes  at various frequencies in the experimental frequency window.
\bibitem{note2} The boson peak is a faster process than the $\beta$ process i.e. at higher frequencies, but they are not temperature dependend and they do not containe a significant information for the state.
\bibitem{note3}  The two functions are not equivalent thus $\beta_{KWW} \neq \beta_{CD}$ for fit of the same data set \cite{lunkeh}.
\end{thebibliography}
\end{document}